\documentclass[%
 aip,
 amsmath,amssymb,
preprint,%
]{revtex4-1}

\usepackage{graphicx}
\usepackage{dcolumn}
\usepackage{bm}
\usepackage{multirow}
\usepackage[utf8]{inputenc}
\usepackage[T1]{fontenc}
\usepackage{mathptmx}
\usepackage{mhchem}
\usepackage{makecell}

\begin{document}

\preprint{AIP/123-QED}

\title[On the happiness of ferroelectric surfaces and its role in water dissociation: the example of bismuth ferrite]
  {On the happiness of ferroelectric surfaces and its role in water dissociation: the example of bismuth ferrite}
%



\author{Ipek Efe}
\author{Nicola A. Spaldin }%
\author{Chiara Gattinoni}
\email{chiara.gattinoni@mat.ethz.ch}
\affiliation{ 
Materials Theory, Department of Materials, ETH Z{\"u}rich, Wolfgang-Pauli-Strasse 27, 8093, Z{\"u}rich, Switzerland}

\date{\today}

\begin{abstract}

We investigate, using density functional theory, how the interaction between the ferroelectric polarization and the chemical structure of the (001) surfaces of bismuth ferrite influences the surface properties and reactivity of this material.
A precise understanding of the surface behavior of ferroelectrics is necessary for their use in surface science applications such as catalysis as well as for their incorporation in microelectronic devices.
Using the (001) surface of bismuth ferrite as a model system we show that the most energetically favoured surface geometries are combinations of surface termination and polarization direction that lead to uncharged, stable surfaces.
On the unfavorable charged surfaces, we explore the compensation mechanisms of surface charges provided by the introduction of point defects and adsorbates, such as water.
Finally, we propose that the special surface properties of bismuth ferrite (001) could be used to produce an effective water splitting cycle through cyclic polarization switching.
\end{abstract}

\maketitle

\vspace{3cm}

The following article has been submitted to the Journal of Chemical Physics.
 
\vspace{1cm}

\section{Introduction}

Transition metal oxides occupy a prominent place in heterogeneous catalysis, and are nowadays the most used industrial catalyst type~\cite{tmoxides_review}.
A variety of industrially relevant processes, for example water splitting or the degradation of pollutant molecules, however, still lack an efficient catalyst.
In the search of novel catalytic materials, the Sabatier principle, which states that effective catalysis occurs when the adsorption between a molecule and a surface is of intermediate strength, is a limiting factor~\cite{vojvodic_nsr_2015}.
However,  the adsorption strength between a molecule and the surface can be controlled by utilizing oxides with tunable functionalities, such as piezo- and ferroelectricity~\cite{kakekhani_jmca_2016}, and research in this field is flourishing~\cite{spaldin_prsa_2020, piezo_medicine_rev, piezoeletricity_review, photocat_fe_10}.
In particular, there is great potential for the use of ferroelectric thin films~\cite{martin_rappe_review} or nanoparticles~\cite{piezoeletricity_review, piezo_medicine_rev, mushtaq_iscience_2018} in electricity generation, water remediation or drug delivery~\cite{spaldin_prsa_2020, photocat_fe_10}.

Ferroelectric materials present a spontaneous switchable bulk polarization, and their surfaces, where reactions occur, are complex. 
In particular, the ferroelectric polarization results in surface bound charges which need to be compensated in order to avoid a polar discontinuity~\cite{stengel_prb_2011}.
Thus, the surface structure of a ferroelectric and, as a consequence, its reactivity are largely determined by the interplay between bound charges and compensation mechanisms~\cite{gattinoni_pnas_2020, levchenko_prl_2008}.

Much progress has been made in our understanding of ferroelectricity at a material's surface.
Indeed, it is now well understood that compensation of the ferroelectric bound charges at a surface occurs preferentially through adsorbates and defect formation rather than by electronic reconstructions~\cite{fong_prl_2006, setvin_science_2018, levchenko_prl_2008, garrity_prb_2013}.
It has also been shown that switching of the surface polarity can be used to promote catalysis for molecular dissociation~\cite{kakekhani_jmca_2016}.
The precise structure of the surface has also been shown to influence the strength and direction of the ferroelectric polarization in thin films, and engineering of surface stoichiometry has been used to manipulate the polarization on ferroelectric surfaces~\cite{wang_prl_2009, highland_prl_2011, saidi_nl_2014, tanase_sr_2016, tian_nc_2018, gattinoni_pnas_2020}.

There are still, however, many open questions regarding the surface science of ferroelectrics~\cite{martin_rappe_review}.
In particular, how the ionic charge in the layers of ferroelectric perovskites interact with the ferroelectric polarization, and the effect of this interplay on the surface structure, is still poorly understood.
Here, we investigate this question in bismuth ferrite (BFO), a material which has a robust ferroelectric polarization at room temperature and, in the (001) direction, neighboring positively-charged Bi$^{3+}$O$^{2-}$ and negatively-charged Fe$^{3+}$O$^{2-}_2$ layers (see Fig.~\ref{fig:unitcell_cartoon_ver8}a).
%
It is also an especially promising catalyst for applications in water remediation~\cite{mushtaq_iscience_2018}, water splitting~\cite{piezoeletricity_review, moniz_nano_2015} and nanoscale drug delivery~\cite{piezo_medicine_rev}.
In the following, we investigate the stability of the (001) surface of BFO, including the interaction of the polarization with defects and water molecule adsorbates.
Our findings allow us to propose a catalytic cycle for efficient water splitting taking advantage of the special properties of BFO (001) surfaces.

\section{Methods}

Density functional theory calculations were performed within the periodic supercell approach using the VASP code~\cite{vasp_1, vasp_2, vasp_3, vasp_4}. 
The optB86b-vdW functional~\cite{klimes}, a revised version of the van der Waals (vdW) density functional of Dion \emph{et al.}~\cite{dion}, was used throughout, as it has been shown to describe well molecular adsorption on transition metal oxides~\cite{vdw1, vdw2, vdw3}. 
Core electrons were replaced by projector augmented wave (PAW) potentials~\cite{paw}, while the valence states (5e$^-$ for Bi, 8e$^-$ for Fe and 6e$^-$ for O) were expanded in plane waves with a cut-off energy of 500 eV. 
In all calculations we used slabs with a $\sqrt{2} a \times \sqrt{2} a$ surface area and $4a$ height (shown in Fig.~\ref{fig:unitcell_cartoon_ver8}a), where $a$ is the lattice parameter of the pseudocubic unit cell.
Using the optB86b-vdW functional the pseudocubic lattice parameter was calculated to be $a = 3.95$ {\AA}, with the $\gamma$ angle in the rhombohedral structure being $\gamma$= 90.23 {$^{\circ}$}.
The difference of the calculated lattice parameters with respect to the experimental structure is below 0.5\%~\cite{kubel_acb_1990}.
A Monkhorst-Pack $k$-point grid of ($5 \times 5 \times 1$) was used for all calculations.
An antiferromagnetic G-type ordering was imposed, which gave a magnetic moment of 4.2 $\mu_{\mathrm{B}}$ per Fe ion in the bulk.
The BFO (001) slabs had a thickness of four cubic unit cells and were separated from their periodic repetitions in the direction perpendicular to the surface by $\sim 20$ \AA\ of vacuum.
Upon testing we found that this thickness was sufficient to converge the adsorption energies of the water molecules (see Table S1).
A dipole correction along the direction perpendicular to the surface was applied, and geometry optimizations were performed with a residual force threshold of 0.01 eV/\AA . 
 %
%
BFO has a large intrinsic polarization, $\mathbf{P}$,  whose experimental value is $\sim 0.9$ C/m$^{2}$ along the (111) direction~\cite{bfo_exp_p}; we calculated P with the formula:
\begin{equation}
\mathbf{P} = \frac{e}{V} \sum_{m=1}^N Q_m \mathbf{u}_m, 
\end{equation}
where $e$ is the charge of the electron, $V$ the unit cell volume, $N$ the number of atoms in the unit cell and $\textbf{u}$ the atomic displacements from the high symmetry positions.
We obtained a value of P=0.86~C/m$^{2}$ when using the formal charges for $Q$ and of P=1.19~C/m$^{2}$ when using the Born effective charges.

Adsorption energies for the water molecules, $\mathrm{E}_{\mathrm{ads}}$, were calculated as:
\begin{equation}
\mathrm{E}_{\mathrm{ads}}=\left(\mathrm{E}_{\mathrm{water/BFO}}-\mathrm{E}_{\mathrm{BFO}} - n \times \mathrm{E}_{\mathrm{water}}   \right)/n,
\label{eq:ads}
\end{equation}
where $\mathrm{E}_{\mathrm{BFO}}$, $\mathrm{E}_{\mathrm{water}}$ and $\mathrm{E}_{\mathrm{water/BFO}}$ are the total energies of the relaxed bare slab, an isolated gas phase water molecule and a system containing $n$ water molecules adsorbed on the slab, respectively. 
Negative values of the adsorption energy indicate favorable (exothermic) adsorption.
Water coverages varying between 1/2 and 1 monolayer (ML) --- where 1 monolayer is one water molecule per surface metal atom --- were considered.

To calculate the charge density differences of Fig.~\ref{fig:happy_intact} we first obtained the real-space charge for the slab/water system ($\rho_{\mathrm{all}}$) and for the isolated slab ($\rho_{\mathrm{slab}}$) and water molecules ($\rho_{\mathrm{water}}$).
The difference was then obtained as:
\begin{equation}
    \rho_{\mathrm{diff}} = \rho_{\mathrm{all}}-\rho_{\mathrm{slab}}-\rho_{\mathrm{water}}.
    \label{eq:rho_diss}
\end{equation}

\section{Results and discussion}

\begin{figure}
\centering
 \includegraphics[width=0.8\textwidth]
 {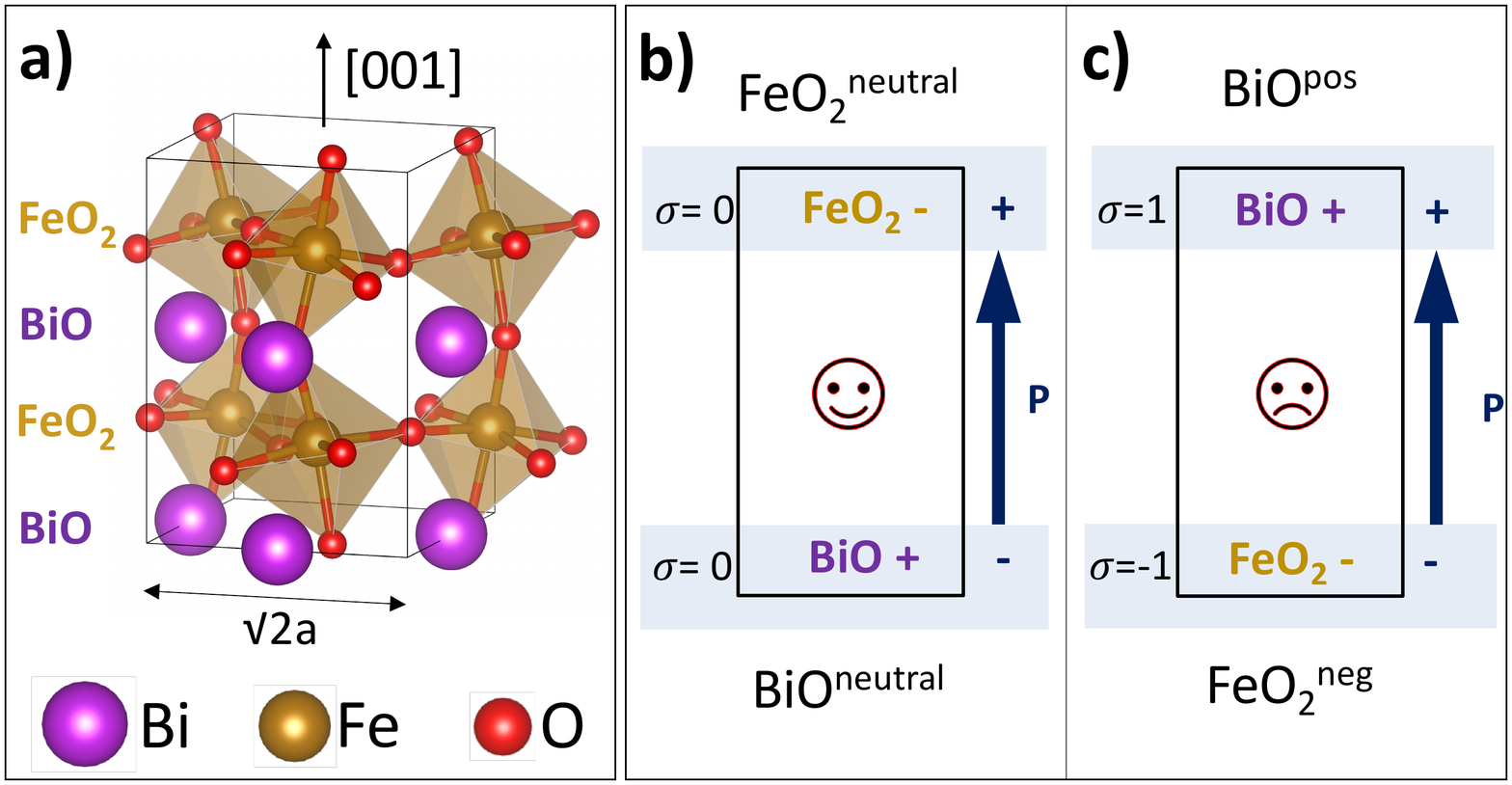}
 \caption{ (a) Unit cell of bismuth ferrite used in this work, with the axes oriented along the $(001)$ and, in plane, the $(110)$ and $(1 \overline{1}0)$ crystallographic directions. The formal charges of the atoms are Bi$^{\mathrm{+3}}$, Fe$^{\mathrm{+3}}$, O$^{\mathrm{-2}}$.  Purple represents Bi, gold Fe and red O atoms. (b) The favorable polarization direction which creates charge-compensated surfaces points from the BiO to the FeO$_2$ termination. (c) The unfavorable polarization direction which creates polar surfaces points from the FeO$_2$ towards the BiO surface termination. $\sigma$ is the surface charge density in C/m$^2$ units.} 
 \label{fig:unitcell_cartoon_ver8}
\end{figure}

BFO (001) has interesting surface properties when we consider the interplay between layer charge and ferroelectric polarization, and they are schematically shown in Fig.~\ref{fig:unitcell_cartoon_ver8}.
The formal charges of Bi$^{3+}$, Fe$^{3+}$ and O$^{2-}$ result, in the (001) direction, in alternating positively charged BiO ($+1$ C/m$^2$) and negatively charged FeO$_2$ ($-1$ C/m$^2$) layers, see Fig.~\ref{fig:unitcell_cartoon_ver8}a.
This surface charge requires a compensating charge of opposite sign and half the magnitude~\cite{hwang_nature, stengel_prb_2011} ($\sim \pm 0.5$ C/m$^2$, negative for BiO and positive for FeO$_2$) to obtain surface stability. 
Remarkably (and coincidentally), the (001) component of the ferroelectric polarization in BiFeO$_3$ has the value $P \sim \pm 0.5$ C/m$^2$ (resulting in the surface charge density of $ \sim \pm 0.5$ C/m$^2$), positive when the polarization is directed towards the surface and negative when away from the surface.
Thus, the interplay of these two contributions of equal magnitude can result either in fully self-compensating surfaces with a total surface charge density $\sigma = 0$ C/m$^2$ in which the surface polarization compensates the layer charge, or highly uncompensated surfaces in which both the layer charge and the surface polarization are contribute to a non-zero surface charge.
\footnote{For a discussion in terms of the Modern Theory of Polarization see Ref.~\citenum{prequel}.}
The self-compensating case, shown in  Fig.~\ref{fig:unitcell_cartoon_ver8}b, occurs in BiO surfaces with the polarization pointing away from them (we will refer to these surfaces as BiO$^{\mathrm{neutral}}$) and FeO$_2$ surfaces with the polarization pointing towards them (FeO$_2^{\mathrm{neutral}}$).
The highly uncompensated surfaces, shown in  Fig.~\ref{fig:unitcell_cartoon_ver8}c, are, instead, the BiO (FeO$_2$) surfaces with the polarization pointing towards (away from) them  and we will refer to these surfaces as BiO$^{\mathrm{pos}}$ and FeO$_2^{\mathrm{neg}}$.

In this work we study the two stoichiometric (001) systems shown in Fig.~\ref{fig:unitcell_cartoon_ver8}b and c.
In panel b there is a fully compensated BFO (001) slab which we we will refer to as the ``happy'' system, since the full surface charge compensation means that there is no polar discontinuity at the surface and the polarization is stable.
The uncompensated slab of panel c will be referred to as the ``unhappy'' system, because the non-zero surface charge density results in an unphysical polar discontinuity, and the surface charge needs to be compensated to render the surface stable~\cite{stengel_prb_2011}.

In the following we explore ways to stabilize the polarization in the unhappy system both with defect engineering and molecular adsorption. 
In particular, we investigate how the different surface electronic properties of the two slabs---their ``happiness'', if you will---affect the geometry and adsorption strength of water.
We show that the resulting polarization-dependent dissociation behavior has great potential for catalytic applications.

\subsection{Achieving surface stability through point defect engineering}

\begin{figure}
\centering
 \includegraphics[trim={0 7cm 0 4cm}, clip, width=0.7\textwidth]{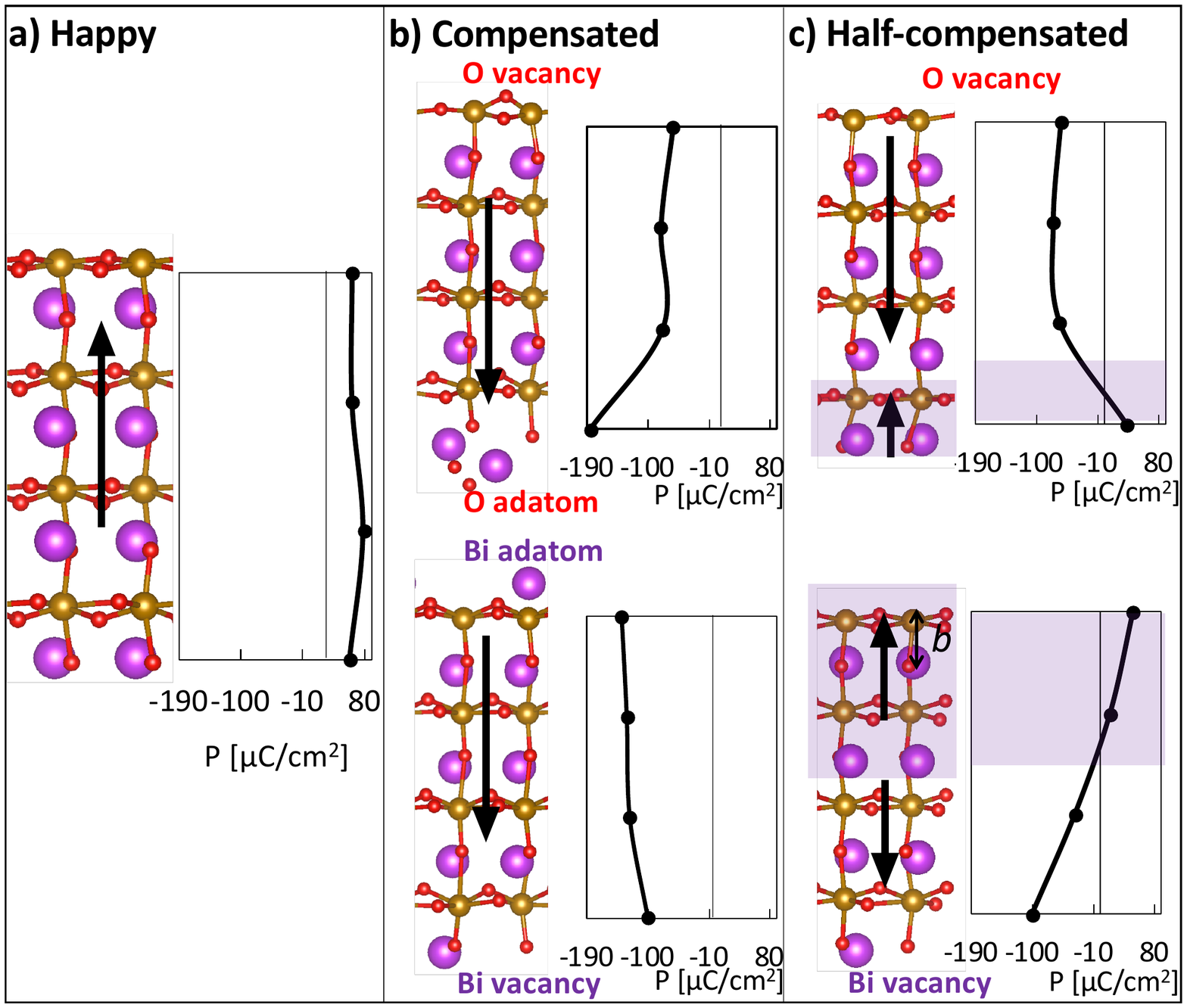}
 \caption{Calculated structures and corresponding layer-by-layer polarization for a range of a four unit cell-thick BFO (001) slabs. a) Happy system. P is constant throughout the slab thickness without further compensation. b) Unhappy slabs with surface charges compensated by O (top) and Bi (bottom) defects. c) unhappy slab with partial compensation of the surface charges. Compensation of the FeO$_2$ surface with an oxygen vacancy (top) and of the BiO surface with a Bi vacancy (bottom). The bond between surface Ti and subsurface O atoms is shown as $b$. The purple shading indicates areas of polarization reversal. The black arrows indicate the polarization direction.}
 \label{fig:stoich_fixed2}
\end{figure}

It is known that point defects and adsorbates can provide charge compensation to ferroelectric surfaces~\cite{gattinoni_pnas_2020, kakekhani_jmca_2016, levchenko_prl_2008}.
As already remarked, the self-compensating surfaces of the happy system have no polar discontinuity and do not require any further compensation.
Indeed, our calculated unit cell by unit cell polarization plotted in Fig.~\ref{fig:stoich_fixed2}a shows that the ferroelectric polarization is stable throughout the slab thickness.
Upon geometry relaxation, the unhappy slab also relaxes into the structure of Fig.~\ref{fig:stoich_fixed2}a, meaning that in order to avoid the polar discontinuity at the surface, the polarization direction reverses, resulting in a happy system.
This indicates that the polarization direction in the unhappy system cannot exist without a means to compensate the surface charges.

To stabilize the unhappy system we consider Bi and O adatoms and vacancies: the positively charged Bi adatom and O vacancy compensate the negatively charged FeO$_2^{\mathrm{neg}}$ surface, the negatively charged Bi vacancy and O adatom compensate the positive BiO$^{\mathrm{pos}}$ surface.
Bi defects, rather than Fe ones, are considered here as they are seen to occur more often in experiments~\cite{bfo_surf_bi}.

Fig.~\ref{fig:stoich_fixed2}b shows the geometry-optimized structure and the unit cell by unit cell polarization in the unhappy slabs compensated with point defects at both surfaces.
We note that, indeed, compensation of the surface charges by vacancies and adatoms is effective in stabilizing the downward-pointing polarization direction in the unhappy slab.
We also observe surface enhancements of the polarization above the bulk value of $90$ $\mu$C/cm$^2$, especially at the surfaces where O and Bi adatoms are present, which are driven by the surface chemistry.
In particular the bonding between a surface Bi and the O adatom (top of Fig.~\ref{fig:stoich_fixed2}b) pulls the Bi atom away from the surface, enhancing the polarization at the BiO surface of the slab.

We also investigated partial compensation of the slab, by including point defects on only one surface, rather than both.
This allows us to understand whether compensation from one surface only is sufficient to ensure stable polarization throughout the slab thickness, and also to separately investigate the BiO$^{\mathrm{pos}}$ and FeO$_2^{\mathrm{neg}}$ surfaces.
The results are shown in Fig.~\ref{fig:stoich_fixed2}c.

On the non-compensated side of the slab polarization reversal occurs,
confirming that compensation on both surfaces is needed to obtain a robust polarization throughout the thickness.
In the uncompensated BiO$^{\mathrm{pos}}$ termination (top of Fig.~\ref{fig:stoich_fixed2}c), polarization reversal occurs only in the outermost BiO surface layer (shown in purple shading), the polarization pointing away from the BiO surface.
For an uncompensated FeO$_2^{\mathrm{neg}}$ surface (bottom of Fig.~\ref{fig:stoich_fixed2}c) the polarization reversal (purple shading) involves the topmost two unit cells and the polarization points towards the surface.
Thus, both uncompensated surfaces become happy by this reversal of the polarization.
The problem of charge compensation now occurs within the slab, where the positive (in the top of Fig.~\ref{fig:stoich_fixed2}c) and the negative (in the bottom of Fig.~\ref{fig:stoich_fixed2}c) ends of the polarization meet creating a polar discontinuity.
Charge compensation in the bulk forces a metallic layer at the site of the polar discontinuity, which requires band bending.
The energy cost of the band bending is however offset by the favorable --- happy --- surface configuration.
However, also the local surface chemistry drives this surface structure.
On BiO$^{\mathrm{pos}}$ the cation has a lone pair of electrons which orients towards the vacuum, pushing the ion towards the subsurface (here, FeO$_2$) layer and creating a ferroelectric polarization pointing away from the surface, and, as a consequence a BiO$^{\mathrm{neutral}}$ surface.
A similar behavior is observed for the PbO surface of lead titanate~\cite{gattinoni_pnas_2020} which also has a lone pair of electrons.
On FeO$_2^{\mathrm{neg}}$ the bond labelled $b$ in Fig.~\ref{fig:stoich_fixed2}c is shorter than in the bulk, as it is generally the case for atomic bonds between the two topmost layers of a slab~\cite{Kern1993_book}. 
This shorter bond $b$ forces the Bi lone pair downwards and the ion upwards, thus imposing a polarization which points towards the surface, which persists, to a lesser degree, in the unit cell below.
Note that in the previous example of lead titanate, no polarization inversion is observed for the TiO$_2$ termination with the polarization pointing away from it~\cite{gattinoni_pnas_2020}.
The difference in behavior between these two ferroelectric perovskites is probably due to the higher relative polarizability of Ti$^{4+}$ compared to Fe$^{3+}$.

Having shown how intrinsic point defects can stabilize the ferroelectric polarization in the unhappy systems, we now investigate how stability can be obtained through adsorbates, by examining the behavior of water on BFO (001).

\subsection{Achieving stability through adsorbates: the example of water}

As well as intrinsic surface defects, adsorbates can play an important role in shaping the surface structure of a ferroelectric~\cite{levchenko_prl_2008}.
The interaction of a surface with water is especially important because of water's ubiquity in air and in solutions, and also because of the potential for applications which arise from the interaction between water and functional materials.
%
%
In the following, we analyse the behavior of water adsorbed on the surfaces of the systems in Fig.~\ref{fig:unitcell_cartoon_ver8}b and c, and reveal how water can stabilize the unhappy system and, in turn, how surface charges affects the water adsorption energy and propensity for dissociation.

\subsubsection{Water adsorption on a happy surface}

\begin{figure}
\centering
 \includegraphics[width=0.6\textwidth, trim={0 5cm 0 5cm}, clip] {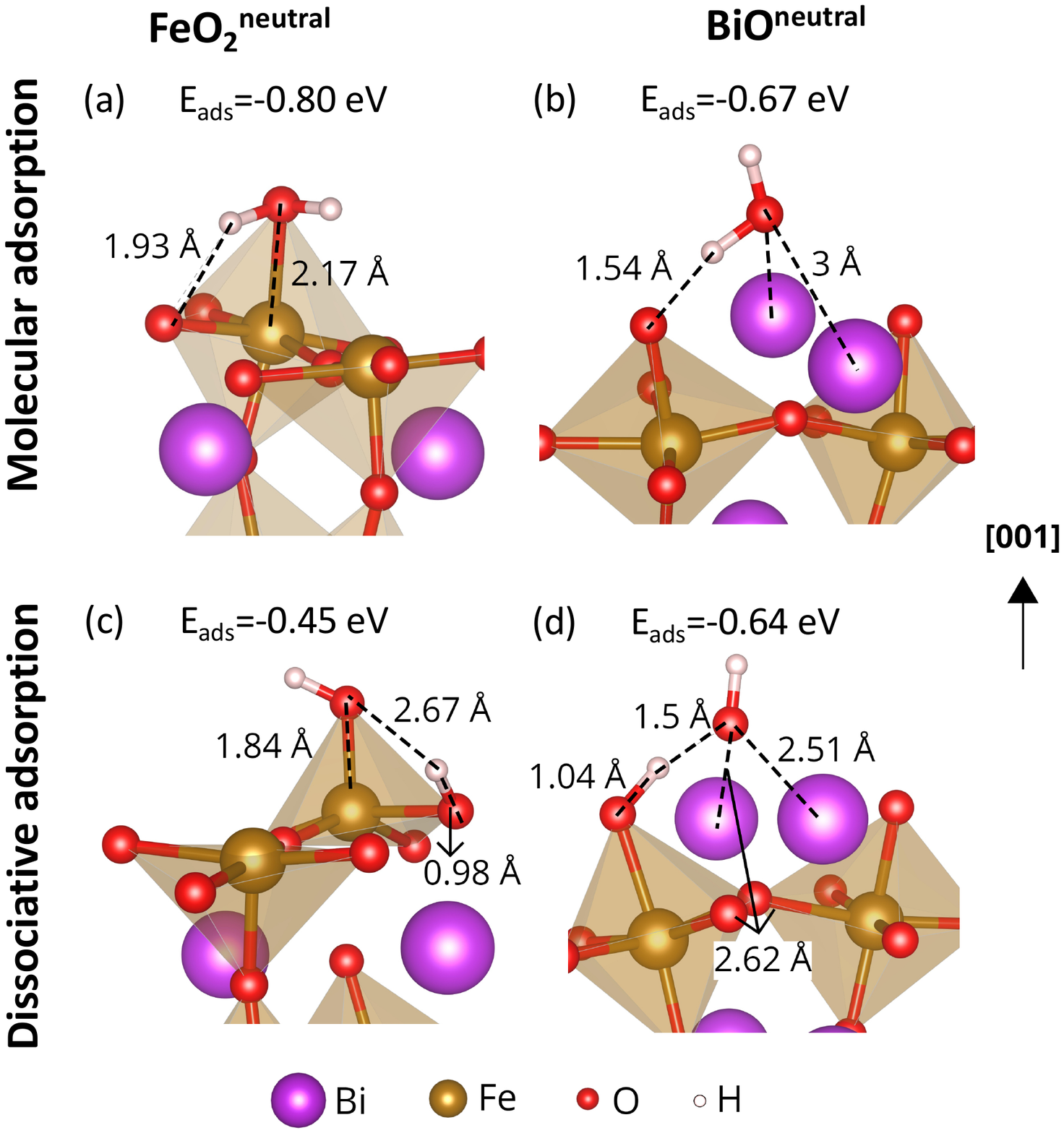}
 \caption{The most favorable adsorption sites for the adsorption of a water molecule, their adsorption energies and the bond distances between the water molecule and the surface ions for the two happy terminations. Molecular adsorption of a water molecule on the (a) FeO$_2^{\mathrm{neutral}}$ and (b) BiO$^{\mathrm{neutral}}$ termination. Dissociative adsorption of a water molecule on (c) FeO$_2^{\mathrm{neutral}}$  and (d) BiO$^{\mathrm{neutral}}$ termination. Purple indicates Bi, gold Fe, red O and white H atoms.  }
 \label{fig:happy_system_ver10}
\end{figure}

\begin{figure} [h]
\centering
 \includegraphics[width=0.70\textwidth, trim={0 7cm 0 0}, clip] {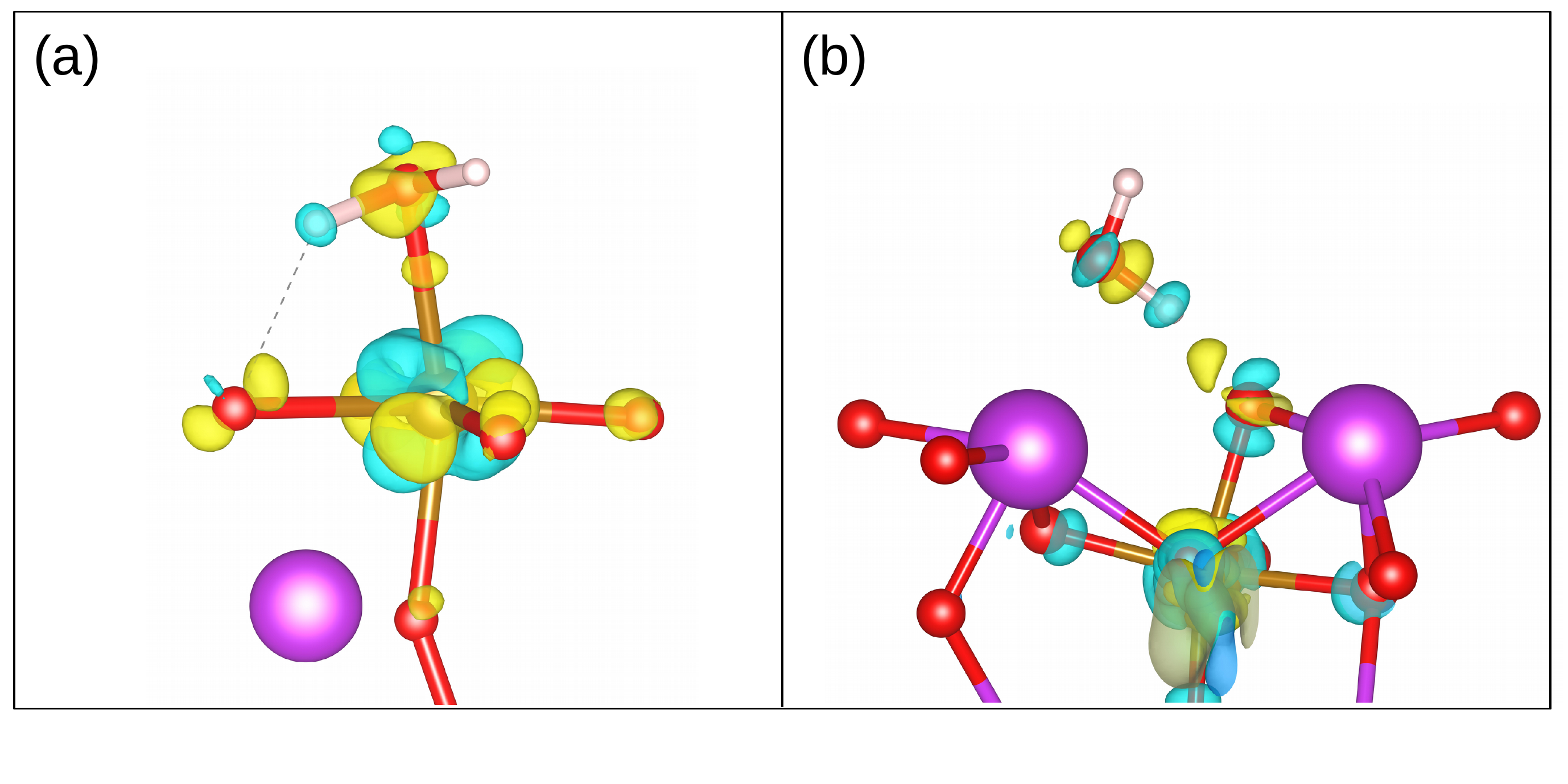}
 \caption{Charge density differences $\rho_{\mathrm{diff}}$, calculated using Eq.~\ref{eq:rho_diss}, for molecular H$_2$O adsorption on the (a) \ce{FeO2} and (b) \ce{BiO} termination. Light blue represents electron density reduction while yellow represents the electron density increase. The isosurface level is 0.01 e/volume.}
 \label{fig:happy_intact}
\end{figure}

We identified the most stable sites for water adsorption on the surfaces of the happy system, and they are shown in Fig.~\ref{fig:happy_system_ver10}.
On FeO$_2^{\mathrm{neutral}}$, the most favorable configuration for molecular H$_2$O adsorption is parallel to the surface with the formation of a 2.17~{\AA} Fe-O bond  and of a 1.93~{\AA} H-O$_{\mathrm{surf}}$ (O$_{\mathrm{surf}}$ is a surface oxygen) bond (see Fig.~\ref{fig:happy_system_ver10}a).
For the BiO$^{\mathrm{neutral}}$ termination, the water O atom sits at the bridging site between two Bi atoms, aligned perpendicularly to the surface.
This configuration permits only one hydrogen bond of length 1.54 {\AA} (Fig.~\ref{fig:happy_system_ver10}b).
Indeed, charge density difference calculations, presented in Fig.~\ref{fig:happy_intact}b, show that minimal charge transfer between the water O and the Bi surface atom occurs.
The water-surface binding is stronger on the FeO$_2^{\mathrm{neutral}}$ termination than on the BiO$^{\mathrm{neutral}}$ by $\sim 0.13$ eV, since in the former molecular adsorption is established by a strong ionic bond and a hydrogen bond (Fig.~\ref{fig:happy_intact}a). 

For a dissociated water molecule, the favored binding sites for the hydroxyl groups are a surface Fe for the FeO$_2^{\mathrm{neutral}}$ termination (see Fig.~\ref{fig:happy_system_ver10}c) and the Bi-Bi bridging site for the BiO$^{\mathrm{neutral}}$ termination (see Fig.~\ref{fig:happy_system_ver10}d), similar configurations to the molecularly adsorbed water.
Also, the rotation of the hydroxyl with respect to the surface is similar to that of the intact water molecule: parallel to the surface on FeO$_2^{\mathrm{neutral}}$ and perpendicular on BiO$^{\mathrm{neutral}}$.
In both cases the H ion binds to an O$_{\mathrm{surf}}$ (Fig.~\ref{fig:happy_system_ver10}c-d).

The adsorption energies in Fig.~\ref{fig:happy_system_ver10} and Table~\ref{table:adsen} show that dissociation of the water molecule is disfavoured on both compensated terminations of a happy BFO (001) slab, by $\sim 30$ meV for BiO$^{\mathrm{neutral}}$ and $\sim 350$ meV for FeO$_2^{\mathrm{neutral}}$. 
%

%

%
%
%
It is worth noting that in all systems the polarization throughout the film is bulk-like and minimally affected by the adsorption of either molecular or dissociated H$_2$O.


\subsubsection{Water adsorption on an unhappy surface}

We next turn our attention to the adsorption of water on the unhappy slab with FeO$_2^{\mathrm{neg}}$ and BiO$^{\mathrm{pos}}$ surfaces, and we find that on this system dissociative water adsorption is favored.

\begin{figure}
\centering
 \includegraphics[width=0.7\textwidth]{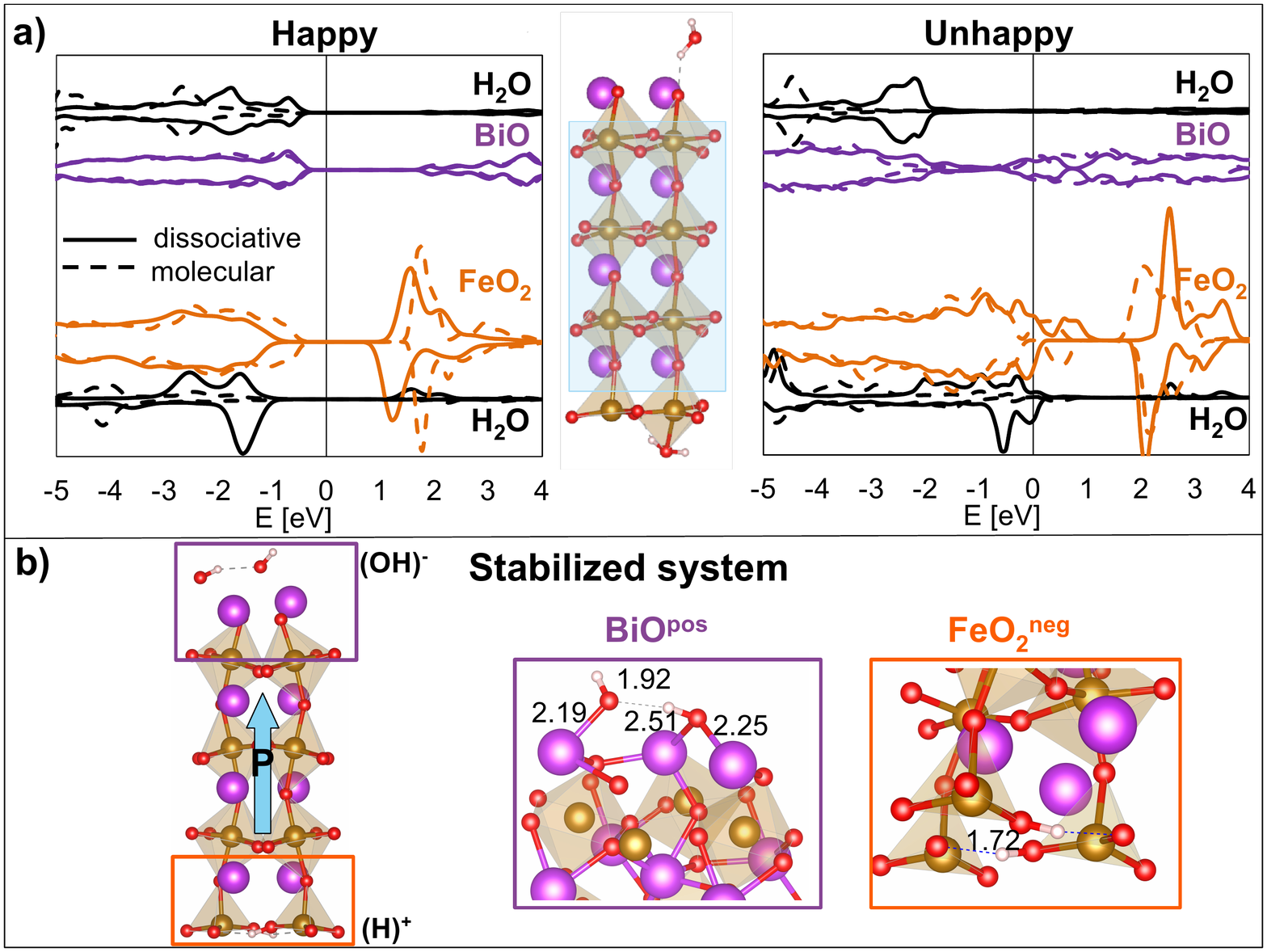}
 \caption{(a) Local density of state ($l$DOS) for the surface layers and water molecules for the "happy" (top) and "unhappy" (bottom) systems. The $l$DOS for water is in black, for the BiO layer in purple and for the FeO$_2$ layer in orange. The black vertical line at E=0 eV is the Fermi level. The dashed (solid) lines are the $l$DOS for molecularly (dissociatively) adsorbed water. A representative system is shown in the middle. The light blue shading indicates the BFO layers which are kept frozen in the calculations of the unhappy systtem. (b) Stabilized system with two water molecules adsorbed dissociatively with OH groups on the BiO termination and H atoms on the FeO$_2$ termination.}
 \label{fig:sad_system}
\end{figure}

Since the unhappy slab is unstable, calculations of the BFO/water system in this section are performed with a ``frozen'' BFO slab: we kept the ionic positions of the inner layers of the slab fixed at the bulk values and allowed only the adsorbed molecules and topmost surface layer, where adsorption occurs, to relax.
The ``frozen'' layers are shown in blue shading in Fig.~\ref{fig:sad_system}a.
%
%
We refer to the adsorption energies with respect to this ``frozen'' substrate as E$_{\mathrm{ads}}^{\mathrm{frozen}}$.

%
We simulated molecularly and dissociatively adsorbed water on the FeO$_2^{\mathrm{neg}}$ and BiO$^{\mathrm{pos}}$ surfaces, and we observed similar adsorption geometries as on the happy slab, both in the preferred adsorption sites and bond lengths (the structures are shown in Fig. S1).
Indeed, on FeO$_2^{\mathrm{neg}}$ the water molecule and the hydroxyl adsorb parallel to the surface; on BiO$^{\mathrm{pos}}$ they adsorb perpendicularly to the surface in the Bi-Bi bridge site.
However, the energy trends are significantly different from the happy case, and dissociative adsorption is favorable on these uncompensated surface terminations.
Indeed, the values in Table~\ref{table:adsen} show that dissociative adosrption is favoured by 290 meV on the FeO$_2^{\mathrm{neg}}$ surface and by 160 meV on BiO$^{\mathrm{pos}}$.

It is worth noting that, despite the similarities in the adsorption geometries, there are significant differences between the electronic structures of the happy and unhappy systems, as highlighted in the local density of state ($l$DOS) of the surface layers and adsorbates in Fig.~\ref{fig:sad_system}a.
In the happy system (left), the $l$DOS of both FeO$_2^{\mathrm{neutral}}$ and BiO$^{\mathrm{neutral}}$ surface layers presents an insulating behavior with a $\sim 2$ eV-wide band gap around the Fermi level, as in bulk BFO.
The conduction and valence band edges on both FeO$_2^{\mathrm{neutral}}$ and BiO$^{\mathrm{neutral}}$ are at the same energy, showing that the ferroelectric polarization is well screened and no band bending occurs through the slab.
Conversely, the FeO$_2^{\mathrm{neg}}$ and BiO$^{\mathrm{pos}}$ surfaces in the unhappy system (right of Fig.~\ref{fig:sad_system}a) are metallic, since the large surface charge of $\pm 1$ C/m$^2$ is compensated by electrons (BiO$^{\mathrm{pos}}$) and holes (FeO$_2^{\mathrm{neg}}$)~\cite{stengel_prb_2011}.

When water is molecularly adsorbed on happy and unhappy surfaces alike (dashed lines in the graphs in Fig.~\ref{fig:sad_system}a), the $l$DOS of the water molecules and of the surface overlap away from the Fermi level and the position of the band edges have only a small effect on the water/surface interaction.
This can explain the similar adsorption energies for intact water on the happy and unhappy system, see Table~\ref{table:adsen}.
The increased stability in the adsorption energy of dissociated water on the unhappy system can instead be related to a partial compensation of the surface charge by the OH$^-$ (on BiO$^{\mathrm{pos}}$) and H$^+$ (on FeO$_2^{\mathrm{neg}}$), which results in change in the $l$DOS at the Fermi level for the unhappy surfaces (especially on FeO$_2^{\mathrm{neg}}$).

\begin{table}[ht] 
\centering 
\begin{tabular}{| c | c | c | c | c| c| c|}
\hline          
\multirow{2}{4em}{System} & \multicolumn{2}{c|}{FeO$_2$} &  \multicolumn{2}{c|}{BiO} & \multicolumn{2}{c|}{Adsorption on both sides} \\ 
 & Intact & Dissociated & Intact & Dissociated & Intact & Dissociated \\ \hline 
Happy & -0.80 & -0.45 & -0.67 & -0.64 & -0.75 & -0.60  \\ \hline
Unhappy (frozen slab) & -0.90 & -1.19  & -0.57 & -0.73 & -0.60 & -0.92  \\ \hline
Stabilized (frozen slab) & --- & --- & --- & --- & --- & -3.16  \\ \hline
Stabilized (relaxed slab) &  --- & --- & --- & --- & --- & -0.47  \\ \hline

\end{tabular}
\caption{Adsorption energies for 0.5 ML of water adsorbed on the BiO and FeO$_2$ surfaces of BFO (001)}
\label{table:adsen}
\end{table}

The molecular adsorption of water on an unhappy slab does not provide adequate charge transfer to stabilize the unfavorable polarization direction, and neither does the co-adsorption of hydroxyl and H on the same surface.
Indeed, for these structures, when we allow the ions in the ``frozen'' slab to relax into their energetically favorable position, we obtain a happy system.
However, stabilization of the polarization in the unhappy system can be achieved when 1 ML of OH$^-$ is adsorbed on the positively charged BiO$^{\mathrm{pos}}$ termination and 1 ML of H$^+$ on the negatively charged FeO$_2^{\mathrm{neg}}$ termination, thus fully compensating the surface charges of $\pm 1$ C/m$^2$.
The adsorption structure is, as expected, at a Bi-Bi bridge site for the hydroxyl groups and atop an O$_{\mathrm{surf}}$ atom for the H atoms.
We will refer to this configuration as the ``stabilized system'' and it is shown in Fig.~\ref{fig:sad_system}b.


The stabilized system is the most stable among the computed water structures on unhappy BFO.
Indeed, the adsorption energy of dissociated water in the stabilized system, with respect to the frozen substrate is E$_{\mathrm{ads}}^{\mathrm{frozen}}=-3.16$ eV/mol, much larger than the values (reported in Table~\ref{table:adsen}) for the structures examined in Fig.~\ref{fig:sad_system}a.

Since the unhappy ferroelectric polarization (from FeO$_2$ to BiO) in the stabilized system in Fig.~\ref{fig:sad_system}b is now fully compensated, we can relax the ionic positions of the whole slab.
We obtain an adsorption energy for the fully relaxed stabilized system (with respect to relaxed happy slab) of E$_{\mathrm{ads}}=-0.47$ eV/mol.
In comparison, water adsorbed on the happy system leads to a more negative (by $0.2$ eV/mol) adsorption energy, see Table~\ref{table:adsen}, and thus to a more energetically favorable structure.
This comparison tells us that the system in Fig.~\ref{fig:sad_system}b, despite being stable, will not occur spontaneously, but will be reached by switching the polarization with an external electric field.

\subsection{Discussion}
The results presented in this work show the complex coupling between the surface chemistry and the ferroelectric polarization at the (001) surfaces of BFO.
The BiO$^{\mathrm{neutral}}$ and FeO$_2^{\mathrm{neutral}}$ terminations of BFO (001) surfaces are charge neutral, thanks to the interaction between layer charge and ferroelectric bound charges.
Upon reversal of the polarization, both surface terminations BiO$^{\mathrm{pos}}$ and FeO$_2^{\mathrm{neg}}$ present a large surface charge that can be effectively compensated by point defects or by dissociated water molecules.
Upon growth of ferroelectric films and nanocrystals, it is very difficult to obtain defect-free surfaces, and these results point to which defects are likely to occur.
Moreover, the surface defect engineering could be important during thin-film growth of unhappy ferroelectric surfaces, where defect formation could be engineered to stabilize or even enhance the surface polarization~\cite{nives_paper, gattinoni_pnas_2020}.

Now we focus on water adsorption and dissociation.
Our calculations reveal that the adsorption mode of water on stoichiometric BFO (001) is highly dependent on the combination of polarization direction and surface termination.
Indeed, we find that compensated surfaces favor the molecular adsorption of water, uncompensated ones dissociative adsorption.
The obtained adsorption structures are in good agreement with previous theoretical work on (001) perovskites, such as strontium ruthenate~\cite{halwidl_2016}, barium hafnate~\cite{bandura_2010}, barium zirconate~\cite{bandura_2010}, strontium titanate ~\cite{guhl_prb_2010} and strontium zirconate ~\cite{evarestov_2007}. 
All these materials are non ferroelectric and have charge-neutral surfaces, therefore, despite similar adsorption structure, we do not necessarily expect the same behavior and the same energy ordering between the intact and dissociated structures.
Indeed, the literature (summarized in Table~II in the SI) shows that water can show a range of different adsorption behavior on the surfaces of these complex materials. 
Together with the surface structure, the interplay of many other factors including neutrality of the crystal ionic layers, lattice parameters~\cite{hu_pccp_2011} and dielectric/metallic characteristics of the materials should be considered to understand the interface chemistry and structure stability. 

\begin{figure}
\centering
 \includegraphics[width=0.99\textwidth] {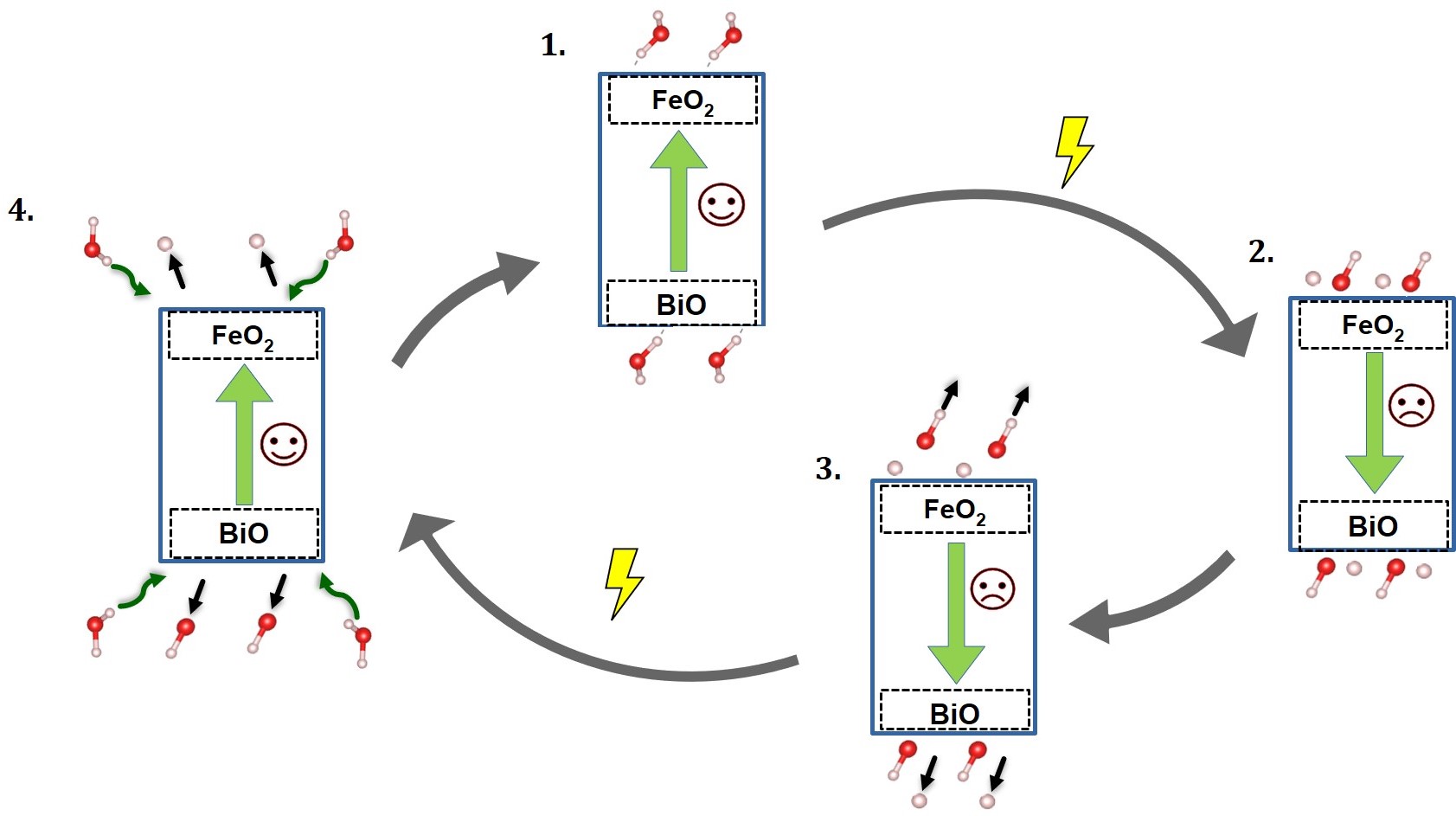}
 \caption{Demonstration of the proposed cyclic process for a water splitting device. (1) Water is adsorbed molecularly on both terminations when the slab is spontaneously polarized in the favorable polarization direction. (2) Switching of the polarization via an external electric field forces the molecularly adsorbed molecules to dissociate on the surface. (3) Selective desorption of the functional groups carrying the same-sign charge with the surfaces. Different functional groups of the water molecules are separated on the opposite surfaces. (4) Desorption of the dissociatively adsorbed groups after switching the polarization back to the favorable direction followed by molecular adsorption of a water molecule.}
 \label{fig:Summary.jpg}
\end{figure}

We believe that this polarization dependence of water dissociation in bismuth ferrite could have interesting ramifications for catalysis.
We propose that the opposite affinity towards water dissociation of the happy and unhappy systems could also be utilized for the creation of a water splitting catalytic cycle, by exploiting the ferro- and piezoelectric properties of BFO as illustrated in Figure~\ref{fig:Summary.jpg}.
The cycle starts with a happy BFO (001) slab which favours molecularly adsorbed water on both BiO$^{\mathrm{neutral}}$ and FeO$_2^{\mathrm{neutral}}$ terminations (Figure~\ref{fig:Summary.jpg}, panel 1). 
Upon switching of the polarization, we obtain an unhappy slab with charged surfaces.
The resulting system favours dissociation of the adsorbed water molecules on BiO$^{\mathrm{pos}}$ and FeO$_2^{\mathrm{neg}}$ surfaces (Figure~\ref{fig:Summary.jpg}, panel 2).
Our calculations then indicate that selective desorption of the H$^{+}$ ion from the BiO$^{\mathrm{pos}}$ termination and OH$^{-}$ group from the FeO$_2^{\mathrm{neg}}$ termination is favorable as it stabilizes the polarization by compensating the surface charges (Figure~\ref{fig:Summary.jpg}, panel 3). 
By further switching of the polarization back into its initial happy direction, competitive adsorption would favor the further removal of dissociation products and the adsorption of molecular H$_2$O (Figure~\ref{fig:Summary.jpg}, panel 4). 
Thus, in principle, cyclical switching of the polarization in a BFO (001) slab immersed in water could efficiently produce H and OH species, which can then by used directly in, \emph{e.g.} the degradation of pollutants~\cite{mushtaq_iscience_2018}, or for H$_2$ production together with a metal cathode.
Polarization switching in nanoscale BFO can be obtained not only with an electric field, but also through mechanical strain~\cite{chen_scirep_2016}.
It could thus be economically achieved with, for example, sound waves~\cite{mushtaq_iscience_2018}.
We hope that this thought experiment can pave the way for the creation of an effective BFO-based water splitting device.

%

\begin{acknowledgments}

C. G. is supported by the European  Union’s  Horizon  2020  research  and  innovation programme under the Marie Sk{\l}odowska-Curie grant agreement No. 744027. N.A.S. acknowledges funding from the European Research Council (ERC) under the European Union’s Horizon 2020 research and innovation programme grant agreement No 810451. I. E. acknowledges the use of the Euler cluster managed by the HPC team at ETH Zurich. C. G.'s work was supported by a grant from the Swiss National Supercomputing Centre (CSCS) under project ID s889.

\end{acknowledgments}

\section*{Data Availability}

The data that supports the findings of this study are available within the article and its supplementary material.

\bibliography{references_bfo_water}

\begin{thebibliography}{45}%
\makeatletter
\providecommand \@ifxundefined [1]{%
 \@ifx{#1\undefined}
}%
\providecommand \@ifnum [1]{%
 \ifnum #1\expandafter \@firstoftwo
 \else \expandafter \@secondoftwo
 \fi
}%
\providecommand \@ifx [1]{%
 \ifx #1\expandafter \@firstoftwo
 \else \expandafter \@secondoftwo
 \fi
}%
\providecommand \natexlab [1]{#1}%
\providecommand \enquote  [1]{``#1''}%
\providecommand \bibnamefont  [1]{#1}%
\providecommand \bibfnamefont [1]{#1}%
\providecommand \citenamefont [1]{#1}%
\providecommand \href@noop [0]{\@secondoftwo}%
\providecommand \href [0]{\begingroup \@sanitize@url \@href}%
\providecommand \@href[1]{\@@startlink{#1}\@@href}%
\providecommand \@@href[1]{\endgroup#1\@@endlink}%
\providecommand \@sanitize@url [0]{\catcode `\\12\catcode `\$12\catcode
  `\&12\catcode `\#12\catcode `\^12\catcode `\_12\catcode `\%12\relax}%
\providecommand \@@startlink[1]{}%
\providecommand \@@endlink[0]{}%
\providecommand \url  [0]{\begingroup\@sanitize@url \@url }%
\providecommand \@url [1]{\endgroup\@href {#1}{\urlprefix }}%
\providecommand \urlprefix  [0]{URL }%
\providecommand \Eprint [0]{\href }%
\providecommand \doibase [0]{http://dx.doi.org/}%
\providecommand \selectlanguage [0]{\@gobble}%
\providecommand \bibinfo  [0]{\@secondoftwo}%
\providecommand \bibfield  [0]{\@secondoftwo}%
\providecommand \translation [1]{[#1]}%
\providecommand \BibitemOpen [0]{}%
\providecommand \bibitemStop [0]{}%
\providecommand \bibitemNoStop [0]{.\EOS\space}%
\providecommand \EOS [0]{\spacefactor3000\relax}%
\providecommand \BibitemShut  [1]{\csname bibitem#1\endcsname}%
\let\auto@bib@innerbib\@empty
\bibitem [{\citenamefont {V{'e}drine}(2017)}]{tmoxides_review}%
  \BibitemOpen
  \bibfield  {author} {\bibinfo {author} {\bibfnamefont {J.}~\bibnamefont
  {V{'e}drine}},\ }\bibfield  {title} {\enquote {\bibinfo {title}
  {Heterogeneous catalysis on metal oxides},}\ }\href@noop {} {\bibfield
  {journal} {\bibinfo  {journal} {Catalysts}\ }\textbf {\bibinfo {volume}
  {7}},\ \bibinfo {pages} {341} (\bibinfo {year} {2017})}\BibitemShut {NoStop}%
\bibitem [{\citenamefont {Vojvodic}\ and\ \citenamefont
  {N\o{}rskov}(2015)}]{vojvodic_nsr_2015}%
  \BibitemOpen
  \bibfield  {author} {\bibinfo {author} {\bibfnamefont {A.}~\bibnamefont
  {Vojvodic}}\ and\ \bibinfo {author} {\bibfnamefont {J.~K.}\ \bibnamefont
  {N\o{}rskov}},\ }\bibfield  {title} {\enquote {\bibinfo {title} {{New design
  paradigm for heterogeneous catalysts}},}\ }\href@noop {} {\bibfield
  {journal} {\bibinfo  {journal} {Nati. Sci. Rev.}\ }\textbf {\bibinfo {volume}
  {2}},\ \bibinfo {pages} {140--143} (\bibinfo {year} {2015})}\BibitemShut
  {NoStop}%
\bibitem [{\citenamefont {Kakekhani}\ and\ \citenamefont
  {Ismail-Beigi}(2016)}]{kakekhani_jmca_2016}%
  \BibitemOpen
  \bibfield  {author} {\bibinfo {author} {\bibfnamefont {A.}~\bibnamefont
  {Kakekhani}}\ and\ \bibinfo {author} {\bibfnamefont {S.}~\bibnamefont
  {Ismail-Beigi}},\ }\bibfield  {title} {\enquote {\bibinfo {title}
  {Ferroelectric oxide surface chemistry: water splitting via
  pyroelectricity},}\ }\href@noop {} {\bibfield  {journal} {\bibinfo  {journal}
  {J. Mater. Chem. A}\ }\textbf {\bibinfo {volume} {4}},\ \bibinfo {pages}
  {5235--5246} (\bibinfo {year} {2016})}\BibitemShut {NoStop}%
\bibitem [{\citenamefont {Spaldin}(2020)}]{spaldin_prsa_2020}%
  \BibitemOpen
  \bibfield  {author} {\bibinfo {author} {\bibfnamefont {N.~A.}\ \bibnamefont
  {Spaldin}},\ }\bibfield  {title} {\enquote {\bibinfo {title} {Multiferroics
  beyond electric-field control of magnetism},}\ }\href@noop {} {\bibfield
  {journal} {\bibinfo  {journal} {Proc. Royal Soc. A}\ }\textbf {\bibinfo
  {volume} {476}},\ \bibinfo {pages} {20190542} (\bibinfo {year}
  {2020})}\BibitemShut {NoStop}%
\bibitem [{\citenamefont {Salim}\ \emph {et~al.}(2018)\citenamefont {Salim},
  \citenamefont {Salim}, \citenamefont {Chandran}, \citenamefont {Aljibori},\
  and\ \citenamefont {Kherbeet}}]{piezo_medicine_rev}%
  \BibitemOpen
  \bibfield  {author} {\bibinfo {author} {\bibfnamefont {M.}~\bibnamefont
  {Salim}}, \bibinfo {author} {\bibfnamefont {D.}~\bibnamefont {Salim}},
  \bibinfo {author} {\bibfnamefont {D.}~\bibnamefont {Chandran}}, \bibinfo
  {author} {\bibfnamefont {H.~S.}\ \bibnamefont {Aljibori}}, \ and\ \bibinfo
  {author} {\bibfnamefont {A.~S.}\ \bibnamefont {Kherbeet}},\ }\bibfield
  {title} {\enquote {\bibinfo {title} {Review of nano piezoelectric devices in
  biomedicine applications},}\ }\href@noop {} {\bibfield  {journal} {\bibinfo
  {journal} {J. Intelligent Mat. Sys. Str.}\ }\textbf {\bibinfo {volume}
  {29}},\ \bibinfo {pages} {2105--2121} (\bibinfo {year} {2018})}\BibitemShut
  {NoStop}%
\bibitem [{\citenamefont {Starr}\ and\ \citenamefont
  {Wang}(2013)}]{piezoeletricity_review}%
  \BibitemOpen
  \bibfield  {author} {\bibinfo {author} {\bibfnamefont {M.~B.}\ \bibnamefont
  {Starr}}\ and\ \bibinfo {author} {\bibfnamefont {X.}~\bibnamefont {Wang}},\
  }\bibfield  {title} {\enquote {\bibinfo {title} {Fundamental analysis of
  piezocatalysis process on the surfaces of strained piezoelectric
  materials},}\ }\href@noop {} {\bibfield  {journal} {\bibinfo  {journal} {Sci.
  Rep.}\ }\textbf {\bibinfo {volume} {3}},\ \bibinfo {pages} {2160} (\bibinfo
  {year} {2013})}\BibitemShut {NoStop}%
\bibitem [{\citenamefont {Bowen}\ \emph {et~al.}(2013)\citenamefont {Bowen},
  \citenamefont {Kim}, \citenamefont {Weaver},\ and\ \citenamefont
  {Dunn}}]{photocat_fe_10}%
  \BibitemOpen
  \bibfield  {author} {\bibinfo {author} {\bibfnamefont {C.~R.}\ \bibnamefont
  {Bowen}}, \bibinfo {author} {\bibfnamefont {H.~A.}\ \bibnamefont {Kim}},
  \bibinfo {author} {\bibfnamefont {P.~M.}\ \bibnamefont {Weaver}}, \ and\
  \bibinfo {author} {\bibfnamefont {S.}~\bibnamefont {Dunn}},\ }\bibfield
  {title} {\enquote {\bibinfo {title} {Piezoelectric and ferroelectric
  materials and structures for energy harvesting applications},}\ }\href@noop
  {} {\bibfield  {journal} {\bibinfo  {journal} {Energy Environ. Sci.}\
  }\textbf {\bibinfo {volume} {7}},\ \bibinfo {pages} {25--44} (\bibinfo {year}
  {2013})}\BibitemShut {NoStop}%
\bibitem [{\citenamefont {Martin}\ and\ \citenamefont
  {Rappe}(2017)}]{martin_rappe_review}%
  \BibitemOpen
  \bibfield  {author} {\bibinfo {author} {\bibfnamefont {L.~W.}\ \bibnamefont
  {Martin}}\ and\ \bibinfo {author} {\bibfnamefont {A.~M.}\ \bibnamefont
  {Rappe}},\ }\bibfield  {title} {\enquote {\bibinfo {title} {Thin-film
  ferroelectric materials and their applications},}\ }\href@noop {} {\bibfield
  {journal} {\bibinfo  {journal} {Nature Rev. Mat.}\ }\textbf {\bibinfo
  {volume} {2}},\ \bibinfo {pages} {16087} (\bibinfo {year}
  {2017})}\BibitemShut {NoStop}%
\bibitem [{\citenamefont {Mushtaq}\ \emph {et~al.}(2018)\citenamefont
  {Mushtaq}, \citenamefont {Chen}, \citenamefont {Hoop}, \citenamefont
  {Torlakcik}, \citenamefont {Pellicer}, \citenamefont {Sort}, \citenamefont
  {Gattinoni}, \citenamefont {Nelson},\ and\ \citenamefont
  {Pan\'{e}}}]{mushtaq_iscience_2018}%
  \BibitemOpen
  \bibfield  {author} {\bibinfo {author} {\bibfnamefont {F.}~\bibnamefont
  {Mushtaq}}, \bibinfo {author} {\bibfnamefont {X.}~\bibnamefont {Chen}},
  \bibinfo {author} {\bibfnamefont {M.}~\bibnamefont {Hoop}}, \bibinfo {author}
  {\bibfnamefont {H.}~\bibnamefont {Torlakcik}}, \bibinfo {author}
  {\bibfnamefont {E.}~\bibnamefont {Pellicer}}, \bibinfo {author}
  {\bibfnamefont {J.}~\bibnamefont {Sort}}, \bibinfo {author} {\bibfnamefont
  {C.}~\bibnamefont {Gattinoni}}, \bibinfo {author} {\bibfnamefont {B.~J.}\
  \bibnamefont {Nelson}}, \ and\ \bibinfo {author} {\bibfnamefont
  {S.}~\bibnamefont {Pan\'{e}}},\ }\bibfield  {title} {\enquote {\bibinfo
  {title} {Piezoelectrically enhanced photocatalysis with {BiFeO}$_3$
  nanostructures for efficient water remediation},}\ }\href@noop {} {\bibfield
  {journal} {\bibinfo  {journal} {iScience}\ }\textbf {\bibinfo {volume} {4}},\
  \bibinfo {pages} {236} (\bibinfo {year} {2018})}\BibitemShut {NoStop}%
\bibitem [{\citenamefont {Stengel}(2011)}]{stengel_prb_2011}%
  \BibitemOpen
  \bibfield  {author} {\bibinfo {author} {\bibfnamefont {M.}~\bibnamefont
  {Stengel}},\ }\bibfield  {title} {\enquote {\bibinfo {title} {Electrostatic
  stability of insulating surfaces: Theory and applications},}\ }\href@noop {}
  {\bibfield  {journal} {\bibinfo  {journal} {Phys. Rev. B}\ }\textbf {\bibinfo
  {volume} {84}},\ \bibinfo {pages} {205432} (\bibinfo {year}
  {2011})}\BibitemShut {NoStop}%
\bibitem [{\citenamefont {Gattinoni}\ \emph {et~al.}(2020)\citenamefont
  {Gattinoni}, \citenamefont {Strkalj}, \citenamefont {H\"ardi}, \citenamefont
  {Fiebig}, \citenamefont {Trassin},\ and\ \citenamefont
  {Spaldin}}]{gattinoni_pnas_2020}%
  \BibitemOpen
  \bibfield  {author} {\bibinfo {author} {\bibfnamefont {C.}~\bibnamefont
  {Gattinoni}}, \bibinfo {author} {\bibfnamefont {N.}~\bibnamefont {Strkalj}},
  \bibinfo {author} {\bibfnamefont {R.}~\bibnamefont {H\"ardi}}, \bibinfo
  {author} {\bibfnamefont {M.}~\bibnamefont {Fiebig}}, \bibinfo {author}
  {\bibfnamefont {M.}~\bibnamefont {Trassin}}, \ and\ \bibinfo {author}
  {\bibfnamefont {N.~A.}\ \bibnamefont {Spaldin}},\ }\bibfield  {title}
  {\enquote {\bibinfo {title} {Interface and surface stabilization of the
  polarization in ferroelectric thin films},}\ }\href@noop {} {\bibfield
  {journal} {\bibinfo  {journal} {Proc. Nat. Ac. Sc.}\ }\textbf {\bibinfo
  {volume} {accepted}},\ \bibinfo {pages} {---} (\bibinfo {year}
  {2020})}\BibitemShut {NoStop}%
\bibitem [{\citenamefont {Levchenko}\ and\ \citenamefont
  {Rappe}(2008)}]{levchenko_prl_2008}%
  \BibitemOpen
  \bibfield  {author} {\bibinfo {author} {\bibfnamefont {S.~V.}\ \bibnamefont
  {Levchenko}}\ and\ \bibinfo {author} {\bibfnamefont {A.~M.}\ \bibnamefont
  {Rappe}},\ }\bibfield  {title} {\enquote {\bibinfo {title} {Influence of
  ferroelectric polarization on the equilibrium stoichiometry of lithium
  niobate (0001) surfaces},}\ }\href@noop {} {\bibfield  {journal} {\bibinfo
  {journal} {Phys. Rev. Lett.}\ }\textbf {\bibinfo {volume} {100}},\ \bibinfo
  {pages} {256101} (\bibinfo {year} {2008})}\BibitemShut {NoStop}%
\bibitem [{\citenamefont {Fong}\ \emph {et~al.}(2006)\citenamefont {Fong},
  \citenamefont {Kolpak}, \citenamefont {Eastman}, \citenamefont {Streiffer},
  \citenamefont {Fuoss}, \citenamefont {Stephenson}, \citenamefont {Thompson},
  \citenamefont {Kim}, \citenamefont {Choi}, \citenamefont {Eom}, \citenamefont
  {Grinberg},\ and\ \citenamefont {Rappe}}]{fong_prl_2006}%
  \BibitemOpen
  \bibfield  {author} {\bibinfo {author} {\bibfnamefont {D.~D.}\ \bibnamefont
  {Fong}}, \bibinfo {author} {\bibfnamefont {A.~M.}\ \bibnamefont {Kolpak}},
  \bibinfo {author} {\bibfnamefont {J.~A.}\ \bibnamefont {Eastman}}, \bibinfo
  {author} {\bibfnamefont {S.~K.}\ \bibnamefont {Streiffer}}, \bibinfo {author}
  {\bibfnamefont {P.~H.}\ \bibnamefont {Fuoss}}, \bibinfo {author}
  {\bibfnamefont {G.~B.}\ \bibnamefont {Stephenson}}, \bibinfo {author}
  {\bibfnamefont {C.}~\bibnamefont {Thompson}}, \bibinfo {author}
  {\bibfnamefont {D.~M.}\ \bibnamefont {Kim}}, \bibinfo {author} {\bibfnamefont
  {K.~J.}\ \bibnamefont {Choi}}, \bibinfo {author} {\bibfnamefont {C.~B.}\
  \bibnamefont {Eom}}, \bibinfo {author} {\bibfnamefont {I.}~\bibnamefont
  {Grinberg}}, \ and\ \bibinfo {author} {\bibfnamefont {A.~M.}\ \bibnamefont
  {Rappe}},\ }\bibfield  {title} {\enquote {\bibinfo {title} {Stabilization of
  monodomain polarization in ultrathin {P}b{T}i{O}$_{3}$ films},}\ }\href@noop
  {} {\bibfield  {journal} {\bibinfo  {journal} {Phys. Rev. Lett.}\ }\textbf
  {\bibinfo {volume} {96}},\ \bibinfo {pages} {127601} (\bibinfo {year}
  {2006})}\BibitemShut {NoStop}%
\bibitem [{\citenamefont {Setvin}\ \emph {et~al.}(2018)\citenamefont {Setvin},
  \citenamefont {Reticcioli}, \citenamefont {Poelzleitner}, \citenamefont
  {Hulva}, \citenamefont {Schmid}, \citenamefont {Boatner}, \citenamefont
  {Franchini},\ and\ \citenamefont {Diebold}}]{setvin_science_2018}%
  \BibitemOpen
  \bibfield  {author} {\bibinfo {author} {\bibfnamefont {M.}~\bibnamefont
  {Setvin}}, \bibinfo {author} {\bibfnamefont {M.}~\bibnamefont {Reticcioli}},
  \bibinfo {author} {\bibfnamefont {F.}~\bibnamefont {Poelzleitner}}, \bibinfo
  {author} {\bibfnamefont {J.}~\bibnamefont {Hulva}}, \bibinfo {author}
  {\bibfnamefont {M.}~\bibnamefont {Schmid}}, \bibinfo {author} {\bibfnamefont
  {L.~A.}\ \bibnamefont {Boatner}}, \bibinfo {author} {\bibfnamefont
  {C.}~\bibnamefont {Franchini}}, \ and\ \bibinfo {author} {\bibfnamefont
  {U.}~\bibnamefont {Diebold}},\ }\bibfield  {title} {\enquote {\bibinfo
  {title} {Polarity compensation mechanisms on the perovskite surface
  {KT}a{O}$_3$(001)},}\ }\href {\doibase 10.1126/science.aar2287} {\bibfield
  {journal} {\bibinfo  {journal} {Science}\ }\textbf {\bibinfo {volume}
  {359}},\ \bibinfo {pages} {572--575} (\bibinfo {year} {2018})}\BibitemShut
  {NoStop}%
\bibitem [{\citenamefont {Garrity}\ \emph {et~al.}(2013)\citenamefont
  {Garrity}, \citenamefont {Kakekhani}, \citenamefont {Kolpak},\ and\
  \citenamefont {Ismail-Beigi}}]{garrity_prb_2013}%
  \BibitemOpen
  \bibfield  {author} {\bibinfo {author} {\bibfnamefont {K.}~\bibnamefont
  {Garrity}}, \bibinfo {author} {\bibfnamefont {A.}~\bibnamefont {Kakekhani}},
  \bibinfo {author} {\bibfnamefont {A.}~\bibnamefont {Kolpak}}, \ and\ \bibinfo
  {author} {\bibfnamefont {S.}~\bibnamefont {Ismail-Beigi}},\ }\bibfield
  {title} {\enquote {\bibinfo {title} {Ferroelectric surface chemistry:
  First-principles study of the {P}b{T}i{O}${}_{3}$ surface},}\ }\href@noop {}
  {\bibfield  {journal} {\bibinfo  {journal} {Phys. Rev. B}\ }\textbf {\bibinfo
  {volume} {88}},\ \bibinfo {pages} {045401} (\bibinfo {year}
  {2013})}\BibitemShut {NoStop}%
\bibitem [{\citenamefont {Wang}\ \emph {et~al.}(2009)\citenamefont {Wang},
  \citenamefont {Fong}, \citenamefont {Jiang}, \citenamefont {Highland},
  \citenamefont {Fuoss}, \citenamefont {Thompson}, \citenamefont {Kolpak},
  \citenamefont {Eastman}, \citenamefont {Streiffer}, \citenamefont {Rappe},\
  and\ \citenamefont {Stephenson}}]{wang_prl_2009}%
  \BibitemOpen
  \bibfield  {author} {\bibinfo {author} {\bibfnamefont {R.~V.}\ \bibnamefont
  {Wang}}, \bibinfo {author} {\bibfnamefont {D.~D.}\ \bibnamefont {Fong}},
  \bibinfo {author} {\bibfnamefont {F.}~\bibnamefont {Jiang}}, \bibinfo
  {author} {\bibfnamefont {M.~J.}\ \bibnamefont {Highland}}, \bibinfo {author}
  {\bibfnamefont {P.~H.}\ \bibnamefont {Fuoss}}, \bibinfo {author}
  {\bibfnamefont {C.}~\bibnamefont {Thompson}}, \bibinfo {author}
  {\bibfnamefont {A.~M.}\ \bibnamefont {Kolpak}}, \bibinfo {author}
  {\bibfnamefont {J.~A.}\ \bibnamefont {Eastman}}, \bibinfo {author}
  {\bibfnamefont {S.~K.}\ \bibnamefont {Streiffer}}, \bibinfo {author}
  {\bibfnamefont {A.~M.}\ \bibnamefont {Rappe}}, \ and\ \bibinfo {author}
  {\bibfnamefont {G.~B.}\ \bibnamefont {Stephenson}},\ }\bibfield  {title}
  {\enquote {\bibinfo {title} {Reversible chemical switching of a ferroelectric
  film},}\ }\href@noop {} {\bibfield  {journal} {\bibinfo  {journal} {Phys.
  Rev. Lett.}\ }\textbf {\bibinfo {volume} {102}},\ \bibinfo {pages} {047601}
  (\bibinfo {year} {2009})}\BibitemShut {NoStop}%
\bibitem [{\citenamefont {Highland}\ \emph {et~al.}(2011)\citenamefont
  {Highland}, \citenamefont {Fister}, \citenamefont {Fong}, \citenamefont
  {Fuoss}, \citenamefont {Thompson}, \citenamefont {Eastman}, \citenamefont
  {Streiffer},\ and\ \citenamefont {Stephenson}}]{highland_prl_2011}%
  \BibitemOpen
  \bibfield  {author} {\bibinfo {author} {\bibfnamefont {M.~J.}\ \bibnamefont
  {Highland}}, \bibinfo {author} {\bibfnamefont {T.~T.}\ \bibnamefont
  {Fister}}, \bibinfo {author} {\bibfnamefont {D.~D.}\ \bibnamefont {Fong}},
  \bibinfo {author} {\bibfnamefont {P.~H.}\ \bibnamefont {Fuoss}}, \bibinfo
  {author} {\bibfnamefont {C.}~\bibnamefont {Thompson}}, \bibinfo {author}
  {\bibfnamefont {J.~A.}\ \bibnamefont {Eastman}}, \bibinfo {author}
  {\bibfnamefont {S.~K.}\ \bibnamefont {Streiffer}}, \ and\ \bibinfo {author}
  {\bibfnamefont {G.~B.}\ \bibnamefont {Stephenson}},\ }\bibfield  {title}
  {\enquote {\bibinfo {title} {Equilibrium polarization of ultrathin
  {P}b{T}i{O}$_{3}$ with surface compensation controlled by oxygen partial
  pressure},}\ }\href@noop {} {\bibfield  {journal} {\bibinfo  {journal} {Phys.
  Rev. Lett.}\ }\textbf {\bibinfo {volume} {107}},\ \bibinfo {pages} {187602}
  (\bibinfo {year} {2011})}\BibitemShut {NoStop}%
\bibitem [{\citenamefont {Saidi}, \citenamefont {Martirez},\ and\ \citenamefont
  {Rappe}(2014)}]{saidi_nl_2014}%
  \BibitemOpen
  \bibfield  {author} {\bibinfo {author} {\bibfnamefont {W.~A.}\ \bibnamefont
  {Saidi}}, \bibinfo {author} {\bibfnamefont {J.~M.~P.}\ \bibnamefont
  {Martirez}}, \ and\ \bibinfo {author} {\bibfnamefont {A.~M.}\ \bibnamefont
  {Rappe}},\ }\bibfield  {title} {\enquote {\bibinfo {title} {Strong reciprocal
  interaction between polarization and surface stoichiometry in oxide
  ferroelectrics},}\ }\href@noop {} {\bibfield  {journal} {\bibinfo  {journal}
  {Nano Lett.}\ }\textbf {\bibinfo {volume} {14}},\ \bibinfo {pages}
  {6711--6717} (\bibinfo {year} {2014})}\BibitemShut {NoStop}%
\bibitem [{\citenamefont {T\v{a}nase}\ \emph {et~al.}(2016)\citenamefont
  {T\v{a}nase}, \citenamefont {Apostol}, \citenamefont {Abramiuc},
  \citenamefont {Tache}, \citenamefont {Hrib}, \citenamefont {Trupin\v{a}},
  \citenamefont {Pintilie},\ and\ \citenamefont {Teodorescu}}]{tanase_sr_2016}%
  \BibitemOpen
  \bibfield  {author} {\bibinfo {author} {\bibfnamefont {L.~C.}\ \bibnamefont
  {T\v{a}nase}}, \bibinfo {author} {\bibfnamefont {N.~G.}\ \bibnamefont
  {Apostol}}, \bibinfo {author} {\bibfnamefont {L.~E.}\ \bibnamefont
  {Abramiuc}}, \bibinfo {author} {\bibfnamefont {C.~A.}\ \bibnamefont {Tache}},
  \bibinfo {author} {\bibfnamefont {L.}~\bibnamefont {Hrib}}, \bibinfo {author}
  {\bibfnamefont {L.}~\bibnamefont {Trupin\v{a}}}, \bibinfo {author}
  {\bibfnamefont {L.}~\bibnamefont {Pintilie}}, \ and\ \bibinfo {author}
  {\bibfnamefont {C.~M.}\ \bibnamefont {Teodorescu}},\ }\bibfield  {title}
  {\enquote {\bibinfo {title} {Ferroelectric triggering of carbon monoxide
  adsorption on lead zirco-titanate (001) surfaces},}\ }\href@noop {}
  {\bibfield  {journal} {\bibinfo  {journal} {Sci. Rep.}\ }\textbf {\bibinfo
  {volume} {6}},\ \bibinfo {pages} {35301} (\bibinfo {year}
  {2016})}\BibitemShut {NoStop}%
\bibitem [{\citenamefont {Tian}\ \emph {et~al.}(2018)\citenamefont {Tian},
  \citenamefont {Wei}, \citenamefont {Zhang}, \citenamefont {Huang},
  \citenamefont {Zhang}, \citenamefont {Zhou}, \citenamefont {Ma},
  \citenamefont {Gu}, \citenamefont {Meng}, \citenamefont {Chen}, \citenamefont
  {Nan},\ and\ \citenamefont {Zhang}}]{tian_nc_2018}%
  \BibitemOpen
  \bibfield  {author} {\bibinfo {author} {\bibfnamefont {Y.}~\bibnamefont
  {Tian}}, \bibinfo {author} {\bibfnamefont {L.}~\bibnamefont {Wei}}, \bibinfo
  {author} {\bibfnamefont {Q.}~\bibnamefont {Zhang}}, \bibinfo {author}
  {\bibfnamefont {H.}~\bibnamefont {Huang}}, \bibinfo {author} {\bibfnamefont
  {Y.}~\bibnamefont {Zhang}}, \bibinfo {author} {\bibfnamefont
  {H.}~\bibnamefont {Zhou}}, \bibinfo {author} {\bibfnamefont {F.}~\bibnamefont
  {Ma}}, \bibinfo {author} {\bibfnamefont {L.}~\bibnamefont {Gu}}, \bibinfo
  {author} {\bibfnamefont {S.}~\bibnamefont {Meng}}, \bibinfo {author}
  {\bibfnamefont {L.-Q.}\ \bibnamefont {Chen}}, \bibinfo {author}
  {\bibfnamefont {C.-W.}\ \bibnamefont {Nan}}, \ and\ \bibinfo {author}
  {\bibfnamefont {J.}~\bibnamefont {Zhang}},\ }\bibfield  {title} {\enquote
  {\bibinfo {title} {Water printing of ferroelectric polarization},}\
  }\href@noop {} {\bibfield  {journal} {\bibinfo  {journal} {Nature Comm.}\
  }\textbf {\bibinfo {volume} {9}},\ \bibinfo {pages} {3809} (\bibinfo {year}
  {2018})}\BibitemShut {NoStop}%
\bibitem [{\citenamefont {Moniz}\ \emph {et~al.}(2015)\citenamefont {Moniz},
  \citenamefont {Blackman}, \citenamefont {Southern}, \citenamefont {Weaver},
  \citenamefont {Tang},\ and\ \citenamefont {Carmalt}}]{moniz_nano_2015}%
  \BibitemOpen
  \bibfield  {author} {\bibinfo {author} {\bibfnamefont {S.~J.~A.}\
  \bibnamefont {Moniz}}, \bibinfo {author} {\bibfnamefont {C.~S.}\ \bibnamefont
  {Blackman}}, \bibinfo {author} {\bibfnamefont {P.}~\bibnamefont {Southern}},
  \bibinfo {author} {\bibfnamefont {P.~M.}\ \bibnamefont {Weaver}}, \bibinfo
  {author} {\bibfnamefont {J.}~\bibnamefont {Tang}}, \ and\ \bibinfo {author}
  {\bibfnamefont {C.~J.}\ \bibnamefont {Carmalt}},\ }\bibfield  {title}
  {\enquote {\bibinfo {title} {Visible-light driven water splitting over
  {BiFeO}$_3$ photoanodes grown via the lpcvd reaction of [{Bi(OtBu)}$_3$] and
  [{Fe(OtBu)}$_3$]$_2$ and enhanced with a surface nickel oxygen evolution
  catalyst},}\ }\href@noop {} {\bibfield  {journal} {\bibinfo  {journal}
  {Nanoscale}\ }\textbf {\bibinfo {volume} {7}},\ \bibinfo {pages}
  {16343--16353} (\bibinfo {year} {2015})}\BibitemShut {NoStop}%
\bibitem [{\citenamefont {Kresse}\ and\ \citenamefont {Hafner}(1993)}]{vasp_1}%
  \BibitemOpen
  \bibfield  {author} {\bibinfo {author} {\bibfnamefont {G.}~\bibnamefont
  {Kresse}}\ and\ \bibinfo {author} {\bibfnamefont {J.}~\bibnamefont
  {Hafner}},\ }\bibfield  {title} {\enquote {\bibinfo {title} {\emph{Ab initio}
  molecular dynamics for liquid metals},}\ }\href@noop {} {\bibfield  {journal}
  {\bibinfo  {journal} {{P}hys. {R}ev. B}\ }\textbf {\bibinfo {volume} {47}},\
  \bibinfo {pages} {558} (\bibinfo {year} {1993})}\BibitemShut {NoStop}%
\bibitem [{\citenamefont {Kresse}\ and\ \citenamefont {Hafner}(1994)}]{vasp_2}%
  \BibitemOpen
  \bibfield  {author} {\bibinfo {author} {\bibfnamefont {G.}~\bibnamefont
  {Kresse}}\ and\ \bibinfo {author} {\bibfnamefont {J.}~\bibnamefont
  {Hafner}},\ }\bibfield  {title} {\enquote {\bibinfo {title} {\emph{Ab initio}
  molecular-dynamics simulation of the liquid-metal–amorphous-semiconductor
  transition in germanium},}\ }\href@noop {} {\bibfield  {journal} {\bibinfo
  {journal} {{P}hys. {R}ev. B}\ }\textbf {\bibinfo {volume} {49}},\ \bibinfo
  {pages} {14251} (\bibinfo {year} {1994})}\BibitemShut {NoStop}%
\bibitem [{\citenamefont {Kresse}\ and\ \citenamefont
  {Furthmueller}(1996{\natexlab{a}})}]{vasp_3}%
  \BibitemOpen
  \bibfield  {author} {\bibinfo {author} {\bibfnamefont {G.}~\bibnamefont
  {Kresse}}\ and\ \bibinfo {author} {\bibfnamefont {J.}~\bibnamefont
  {Furthmueller}},\ }\bibfield  {title} {\enquote {\bibinfo {title} {Efficiency
  of ab-initio total energy calculations for metals and semiconductors using a
  plane-wave basis set},}\ }\href@noop {} {\bibfield  {journal} {\bibinfo
  {journal} {{C}omp. {M}at. {S}ci.}\ }\textbf {\bibinfo {volume} {6}},\
  \bibinfo {pages} {15} (\bibinfo {year} {1996}{\natexlab{a}})}\BibitemShut
  {NoStop}%
\bibitem [{\citenamefont {Kresse}\ and\ \citenamefont
  {Furthmueller}(1996{\natexlab{b}})}]{vasp_4}%
  \BibitemOpen
  \bibfield  {author} {\bibinfo {author} {\bibfnamefont {G.}~\bibnamefont
  {Kresse}}\ and\ \bibinfo {author} {\bibfnamefont {J.}~\bibnamefont
  {Furthmueller}},\ }\bibfield  {title} {\enquote {\bibinfo {title} {Efficient
  iterative schemes for \emph{ab initio} total-energy calculations using a
  plane-wave basis set},}\ }\href@noop {} {\bibfield  {journal} {\bibinfo
  {journal} {{P}hys. {R}ev. B}\ }\textbf {\bibinfo {volume} {54}},\ \bibinfo
  {pages} {11169} (\bibinfo {year} {1996}{\natexlab{b}})}\BibitemShut {NoStop}%
\bibitem [{\citenamefont {Klime\v{s}}, \citenamefont {Bowler},\ and\
  \citenamefont {Michaelides}(2011)}]{klimes}%
  \BibitemOpen
  \bibfield  {author} {\bibinfo {author} {\bibfnamefont {J.}~\bibnamefont
  {Klime\v{s}}}, \bibinfo {author} {\bibfnamefont {D.~R.}\ \bibnamefont
  {Bowler}}, \ and\ \bibinfo {author} {\bibfnamefont {A.}~\bibnamefont
  {Michaelides}},\ }\bibfield  {title} {\enquote {\bibinfo {title} {Van der
  {W}aals density functionals applied to solids},}\ }\href@noop {} {\bibfield
  {journal} {\bibinfo  {journal} {Phys. Rev. B}\ }\textbf {\bibinfo {volume}
  {83}},\ \bibinfo {pages} {195131} (\bibinfo {year} {2011})}\BibitemShut
  {NoStop}%
\bibitem [{\citenamefont {Dion}\ \emph {et~al.}(2004)\citenamefont {Dion},
  \citenamefont {Rydberg}, \citenamefont {Schröder}, \citenamefont
  {Langreth},\ and\ \citenamefont {Lundqvist}}]{dion}%
  \BibitemOpen
  \bibfield  {author} {\bibinfo {author} {\bibfnamefont {M.}~\bibnamefont
  {Dion}}, \bibinfo {author} {\bibfnamefont {H.}~\bibnamefont {Rydberg}},
  \bibinfo {author} {\bibfnamefont {E.}~\bibnamefont {Schröder}}, \bibinfo
  {author} {\bibfnamefont {D.~C.}\ \bibnamefont {Langreth}}, \ and\ \bibinfo
  {author} {\bibfnamefont {B.~I.}\ \bibnamefont {Lundqvist}},\ }\bibfield
  {title} {\enquote {\bibinfo {title} {Van der {W}aals density functional for
  general geometries},}\ }\href@noop {} {\bibfield  {journal} {\bibinfo
  {journal} {Phys. Rev. Lett.}\ }\textbf {\bibinfo {volume} {92}},\ \bibinfo
  {pages} {246401} (\bibinfo {year} {2004})}\BibitemShut {NoStop}%
\bibitem [{\citenamefont {Dabaghmanesh}, \citenamefont {Neyts},\ and\
  \citenamefont {Partoens}(2016)}]{vdw1}%
  \BibitemOpen
  \bibfield  {author} {\bibinfo {author} {\bibfnamefont {S.}~\bibnamefont
  {Dabaghmanesh}}, \bibinfo {author} {\bibfnamefont {E.~C.}\ \bibnamefont
  {Neyts}}, \ and\ \bibinfo {author} {\bibfnamefont {B.}~\bibnamefont
  {Partoens}},\ }\bibfield  {title} {\enquote {\bibinfo {title} {Van der waals
  density functionals applied to corundum-type sesquioxides: bulk properties
  and adsorption of {CH$_3$} and {C$_6$H$_6$} on (0001) surfaces},}\
  }\href@noop {} {\bibfield  {journal} {\bibinfo  {journal} {Phys. Chem. Chem.
  Phys.}\ }\textbf {\bibinfo {volume} {18}},\ \bibinfo {pages} {23139--23146}
  (\bibinfo {year} {2016})}\BibitemShut {NoStop}%
\bibitem [{\citenamefont {Roy}, \citenamefont {Kar},\ and\ \citenamefont
  {Leszczynski}(2018)}]{vdw2}%
  \BibitemOpen
  \bibfield  {author} {\bibinfo {author} {\bibfnamefont {J.}~\bibnamefont
  {Roy}}, \bibinfo {author} {\bibfnamefont {S.}~\bibnamefont {Kar}}, \ and\
  \bibinfo {author} {\bibfnamefont {J.}~\bibnamefont {Leszczynski}},\
  }\bibfield  {title} {\enquote {\bibinfo {title} {Insight into the
  optoelectronic properties of designed solar cells efficient
  tetrahydroquinoline dye-sensitizers on {T}i{O}$_2$(101) surface: first
  principles approach},}\ }\href@noop {} {\bibfield  {journal} {\bibinfo
  {journal} {Sci. Rep.}\ }\textbf {\bibinfo {volume} {8}},\ \bibinfo {pages}
  {10997} (\bibinfo {year} {2018})}\BibitemShut {NoStop}%
\bibitem [{\citenamefont {Lozano}\ \emph {et~al.}(2017)\citenamefont {Lozano},
  \citenamefont {Escribano}, \citenamefont {Akhmatskaya},\ and\ \citenamefont
  {Carrasco}}]{vdw3}%
  \BibitemOpen
  \bibfield  {author} {\bibinfo {author} {\bibfnamefont {A.}~\bibnamefont
  {Lozano}}, \bibinfo {author} {\bibfnamefont {B.}~\bibnamefont {Escribano}},
  \bibinfo {author} {\bibfnamefont {E.}~\bibnamefont {Akhmatskaya}}, \ and\
  \bibinfo {author} {\bibfnamefont {J.}~\bibnamefont {Carrasco}},\ }\bibfield
  {title} {\enquote {\bibinfo {title} {Assessment of van der {W}aals inclusive
  density functional theory methods for layered electroactive materials},}\
  }\href@noop {} {\bibfield  {journal} {\bibinfo  {journal} {Phys. Chem. Chem.
  Phys.}\ }\textbf {\bibinfo {volume} {19}},\ \bibinfo {pages} {10133--10139}
  (\bibinfo {year} {2017})}\BibitemShut {NoStop}%
\bibitem [{\citenamefont {Kresse}\ and\ \citenamefont {Joubert}(1999)}]{paw}%
  \BibitemOpen
  \bibfield  {author} {\bibinfo {author} {\bibfnamefont {G.}~\bibnamefont
  {Kresse}}\ and\ \bibinfo {author} {\bibfnamefont {D.}~\bibnamefont
  {Joubert}},\ }\bibfield  {title} {\enquote {\bibinfo {title} {{From ultrasoft
  pseudopotentials to the projector augmented-wave method}},}\ }\href@noop {}
  {\bibfield  {journal} {\bibinfo  {journal} {Phys. Rev. B}\ }\textbf {\bibinfo
  {volume} {{59}}},\ \bibinfo {pages} {{1758}} (\bibinfo {year}
  {{1999}})}\BibitemShut {NoStop}%
\bibitem [{\citenamefont {Kubel}\ and\ \citenamefont
  {Schmid}(1990)}]{kubel_acb_1990}%
  \BibitemOpen
  \bibfield  {author} {\bibinfo {author} {\bibfnamefont {F.}~\bibnamefont
  {Kubel}}\ and\ \bibinfo {author} {\bibfnamefont {H.}~\bibnamefont {Schmid}},\
  }\bibfield  {title} {\enquote {\bibinfo {title} {{Structure of a
  ferroelectric and ferroelastic monodomain crystal of the perovskite
  BiFeO${\sb 3}$}},}\ }\href@noop {} {\bibfield  {journal} {\bibinfo  {journal}
  {Acta Crys. B}\ }\textbf {\bibinfo {volume} {46}},\ \bibinfo {pages}
  {698--702} (\bibinfo {year} {1990})}\BibitemShut {NoStop}%
\bibitem [{\citenamefont {Rojac}\ \emph {et~al.}(2018)\citenamefont {Rojac},
  \citenamefont {Khomyakova}, \citenamefont {Walker}, \citenamefont {Ursic},\
  and\ \citenamefont {Bencan}}]{bfo_exp_p}%
  \BibitemOpen
  \bibfield  {author} {\bibinfo {author} {\bibfnamefont {T.}~\bibnamefont
  {Rojac}}, \bibinfo {author} {\bibfnamefont {E.}~\bibnamefont {Khomyakova}},
  \bibinfo {author} {\bibfnamefont {J.}~\bibnamefont {Walker}}, \bibinfo
  {author} {\bibfnamefont {H.}~\bibnamefont {Ursic}}, \ and\ \bibinfo {author}
  {\bibfnamefont {A.}~\bibnamefont {Bencan}},\ }\bibfield  {title} {\enquote
  {\bibinfo {title} {{BiFeO$_3$} ceramics and thick films: Processing issues
  and electromechanical properties},}\ }in\ \href@noop {} {\emph {\bibinfo
  {booktitle} {Magnetic, Ferroelectric, and Multiferroic Metal Oxides}}},\
  \bibinfo {series and number} {Metal Oxides},\ \bibinfo {editor} {edited by\
  \bibinfo {editor} {\bibfnamefont {B.~D.}\ \bibnamefont {Stojanovic}}}\
  (\bibinfo  {publisher} {Elsevier},\ \bibinfo {year} {2018})\ pp.\ \bibinfo
  {pages} {515 -- 525}\BibitemShut {NoStop}%
\bibitem [{\citenamefont {Ohtomo}\ and\ \citenamefont
  {Hwang}(2004)}]{hwang_nature}%
  \BibitemOpen
  \bibfield  {author} {\bibinfo {author} {\bibfnamefont {A.}~\bibnamefont
  {Ohtomo}}\ and\ \bibinfo {author} {\bibfnamefont {H.~Y.}\ \bibnamefont
  {Hwang}},\ }\bibfield  {title} {\enquote {\bibinfo {title} {A high-mobility
  electron gas at the {LaAlO}$_3$/{SrTiO}$_3$ heterointerface},}\ }\href@noop
  {} {\bibfield  {journal} {\bibinfo  {journal} {Nature}\ }\textbf {\bibinfo
  {volume} {427}},\ \bibinfo {pages} {423--426} (\bibinfo {year}
  {2004})}\BibitemShut {NoStop}%
\bibitem [{Note1()}]{Note1}%
  \BibitemOpen
  \bibinfo {note} {For a discussion in terms of the Modern Theory of
  Polarization see Ref.~\protect \citenum {prequel}.}\BibitemShut {Stop}%
\bibitem [{\citenamefont {Jin}\ \emph {et~al.}(2017)\citenamefont {Jin},
  \citenamefont {Xu}, \citenamefont {Zeng}, \citenamefont {Lu}, \citenamefont
  {Barthel}, \citenamefont {Schulthess}, \citenamefont {Dunin-Borkowski},
  \citenamefont {Wang},\ and\ \citenamefont {Jia}}]{bfo_surf_bi}%
  \BibitemOpen
  \bibfield  {author} {\bibinfo {author} {\bibfnamefont {L.}~\bibnamefont
  {Jin}}, \bibinfo {author} {\bibfnamefont {P.~X.}\ \bibnamefont {Xu}},
  \bibinfo {author} {\bibfnamefont {Y.}~\bibnamefont {Zeng}}, \bibinfo {author}
  {\bibfnamefont {L.}~\bibnamefont {Lu}}, \bibinfo {author} {\bibfnamefont
  {J.}~\bibnamefont {Barthel}}, \bibinfo {author} {\bibfnamefont
  {T.}~\bibnamefont {Schulthess}}, \bibinfo {author} {\bibfnamefont {R.~E.}\
  \bibnamefont {Dunin-Borkowski}}, \bibinfo {author} {\bibfnamefont
  {H.}~\bibnamefont {Wang}}, \ and\ \bibinfo {author} {\bibfnamefont {C.~L.}\
  \bibnamefont {Jia}},\ }\bibfield  {title} {\enquote {\bibinfo {title}
  {Surface reconstructions and related local properties of a {BiFeO}$_3$ thin
  film},}\ }\href@noop {} {\bibfield  {journal} {\bibinfo  {journal}
  {Scientific reports}\ }\textbf {\bibinfo {volume} {7}},\ \bibinfo {pages}
  {39698} (\bibinfo {year} {2017})}\BibitemShut {NoStop}%
\bibitem [{\citenamefont {Kern}(1993)}]{Kern1993_book}%
  \BibitemOpen
  \bibfield  {author} {\bibinfo {author} {\bibfnamefont {K.}~\bibnamefont
  {Kern}},\ }\enquote {\bibinfo {title} {Restructuring at surfaces},}\ in\
  \href@noop {} {\emph {\bibinfo {booktitle} {Surface Science: Principles and
  Applications}}},\ \bibinfo {editor} {edited by\ \bibinfo {editor}
  {\bibfnamefont {R.~F.}\ \bibnamefont {Howe}}, \bibinfo {editor}
  {\bibfnamefont {R.~N.}\ \bibnamefont {Lamb}}, \ and\ \bibinfo {editor}
  {\bibfnamefont {K.}~\bibnamefont {Wandelt}}}\ (\bibinfo  {publisher}
  {Springer Berlin Heidelberg},\ \bibinfo {address} {Berlin, Heidelberg},\
  \bibinfo {year} {1993})\ pp.\ \bibinfo {pages} {81--94}\BibitemShut {NoStop}%
\bibitem [{\citenamefont {Strkalj}\ \emph {et~al.}(2020)\citenamefont
  {Strkalj}, \citenamefont {Gattinoni}, \citenamefont {Vogel}, \citenamefont
  {Campanini}, \citenamefont {Haerdi}, \citenamefont {Rossi}, \citenamefont
  {Rossell}, \citenamefont {Spaldin}, \citenamefont {Fiebig},\ and\
  \citenamefont {Trassin}}]{nives_paper}%
  \BibitemOpen
  \bibfield  {author} {\bibinfo {author} {\bibfnamefont {N.}~\bibnamefont
  {Strkalj}}, \bibinfo {author} {\bibfnamefont {C.}~\bibnamefont {Gattinoni}},
  \bibinfo {author} {\bibfnamefont {A.}~\bibnamefont {Vogel}}, \bibinfo
  {author} {\bibfnamefont {M.}~\bibnamefont {Campanini}}, \bibinfo {author}
  {\bibfnamefont {R.}~\bibnamefont {Haerdi}}, \bibinfo {author} {\bibfnamefont
  {A.}~\bibnamefont {Rossi}}, \bibinfo {author} {\bibfnamefont {M.~D.}\
  \bibnamefont {Rossell}}, \bibinfo {author} {\bibfnamefont {N.~A.}\
  \bibnamefont {Spaldin}}, \bibinfo {author} {\bibfnamefont {M.}~\bibnamefont
  {Fiebig}}, \ and\ \bibinfo {author} {\bibfnamefont {M.}~\bibnamefont
  {Trassin}},\ }\bibfield  {title} {\enquote {\bibinfo {title} {Bilateral
  interface control of nanoscale ferroelectricity},}\ }\href@noop {} {\bibfield
   {journal} {\bibinfo  {journal} {in preparation}\ } (\bibinfo {year}
  {2020})}\BibitemShut {NoStop}%
\bibitem [{\citenamefont {Halwidl}\ \emph {et~al.}(2016)\citenamefont
  {Halwidl}, \citenamefont {Stöger}, \citenamefont {Mayr-Schmölzer},
  \citenamefont {Pavelec}, \citenamefont {Fobes}, \citenamefont {Peng},
  \citenamefont {Mao}, \citenamefont {Parkinson}, \citenamefont {Schmid},
  \citenamefont {Mittendorfer}, \citenamefont {Redinger},\ and\ \citenamefont
  {Diebold}}]{halwidl_2016}%
  \BibitemOpen
  \bibfield  {author} {\bibinfo {author} {\bibfnamefont {D.}~\bibnamefont
  {Halwidl}}, \bibinfo {author} {\bibfnamefont {B.}~\bibnamefont {Stöger}},
  \bibinfo {author} {\bibfnamefont {W.}~\bibnamefont {Mayr-Schmölzer}},
  \bibinfo {author} {\bibfnamefont {J.}~\bibnamefont {Pavelec}}, \bibinfo
  {author} {\bibfnamefont {D.}~\bibnamefont {Fobes}}, \bibinfo {author}
  {\bibfnamefont {J.}~\bibnamefont {Peng}}, \bibinfo {author} {\bibfnamefont
  {Z.}~\bibnamefont {Mao}}, \bibinfo {author} {\bibfnamefont {G.~S.}\
  \bibnamefont {Parkinson}}, \bibinfo {author} {\bibfnamefont {M.}~\bibnamefont
  {Schmid}}, \bibinfo {author} {\bibfnamefont {F.}~\bibnamefont
  {Mittendorfer}}, \bibinfo {author} {\bibfnamefont {J.}~\bibnamefont
  {Redinger}}, \ and\ \bibinfo {author} {\bibfnamefont {U.}~\bibnamefont
  {Diebold}},\ }\bibfield  {title} {\enquote {\bibinfo {title} {Adsorption of
  water at the {SrO} surface of ruthenates},}\ }\href@noop {} {\bibfield
  {journal} {\bibinfo  {journal} {Nature Mat.}\ }\textbf {\bibinfo {volume}
  {15}},\ \bibinfo {pages} {450--455} (\bibinfo {year} {2016})}\BibitemShut
  {NoStop}%
\bibitem [{\citenamefont {Bandura}, \citenamefont {Evarestov},\ and\
  \citenamefont {Kuruch}(2010)}]{bandura_2010}%
  \BibitemOpen
  \bibfield  {author} {\bibinfo {author} {\bibfnamefont {A.~V.}\ \bibnamefont
  {Bandura}}, \bibinfo {author} {\bibfnamefont {R.~A.}\ \bibnamefont
  {Evarestov}}, \ and\ \bibinfo {author} {\bibfnamefont {D.~D.}\ \bibnamefont
  {Kuruch}},\ }\bibfield  {title} {\enquote {\bibinfo {title} {Hybrid
  {HF}-{DFT} modeling of monolayer water adsorption on (001) surface of cubic
  {BaHfO$_3$} and {BaZrO$_3$} crystals},}\ }\href@noop {} {\bibfield  {journal}
  {\bibinfo  {journal} {Surf. Sci.}\ }\textbf {\bibinfo {volume} {604}},\
  \bibinfo {pages} {1591--1597} (\bibinfo {year} {2010})}\BibitemShut {NoStop}%
\bibitem [{\citenamefont {Guhl}, \citenamefont {Miller},\ and\ \citenamefont
  {Reuter}(2010)}]{guhl_prb_2010}%
  \BibitemOpen
  \bibfield  {author} {\bibinfo {author} {\bibfnamefont {H.}~\bibnamefont
  {Guhl}}, \bibinfo {author} {\bibfnamefont {W.}~\bibnamefont {Miller}}, \ and\
  \bibinfo {author} {\bibfnamefont {K.}~\bibnamefont {Reuter}},\ }\bibfield
  {title} {\enquote {\bibinfo {title} {Water adsorption and dissociation on
  {SrTiO}$_{3}(001)$ revisited: A density functional theory study},}\ }\href
  {\doibase 10.1103/PhysRevB.81.155455} {\bibfield  {journal} {\bibinfo
  {journal} {Phys. Rev. B}\ }\textbf {\bibinfo {volume} {81}},\ \bibinfo
  {pages} {155455} (\bibinfo {year} {2010})}\BibitemShut {NoStop}%
\bibitem [{\citenamefont {Evarestov}, \citenamefont {Bandura},\ and\
  \citenamefont {Alexandrov}(2007)}]{evarestov_2007}%
  \BibitemOpen
  \bibfield  {author} {\bibinfo {author} {\bibfnamefont {R.~A.}\ \bibnamefont
  {Evarestov}}, \bibinfo {author} {\bibfnamefont {A.~V.}\ \bibnamefont
  {Bandura}}, \ and\ \bibinfo {author} {\bibfnamefont {V.~E.}\ \bibnamefont
  {Alexandrov}},\ }\bibfield  {title} {\enquote {\bibinfo {title} {Adsorption
  of water on (001) surface of {SrTiO$_3$} and {SrZrO$_3$} cubic perovskites:
  {Hybrid} {HF}-{DFT} {LCAO} calculations},}\ }\href@noop {} {\bibfield
  {journal} {\bibinfo  {journal} {Surf. Sci.}\ }\textbf {\bibinfo {volume}
  {601}},\ \bibinfo {pages} {1844--1856} (\bibinfo {year} {2007})}\BibitemShut
  {NoStop}%
\bibitem [{\citenamefont {Hu}\ \emph {et~al.}(2011)\citenamefont {Hu},
  \citenamefont {Carrasco}, \citenamefont {Klimeš},\ and\ \citenamefont
  {Michaelides}}]{hu_pccp_2011}%
  \BibitemOpen
  \bibfield  {author} {\bibinfo {author} {\bibfnamefont {X.~L.}\ \bibnamefont
  {Hu}}, \bibinfo {author} {\bibfnamefont {J.}~\bibnamefont {Carrasco}},
  \bibinfo {author} {\bibfnamefont {J.}~\bibnamefont {Klimeš}}, \ and\
  \bibinfo {author} {\bibfnamefont {A.}~\bibnamefont {Michaelides}},\
  }\bibfield  {title} {\enquote {\bibinfo {title} {Trends in water monomer
  adsorption and dissociation on flat insulating surfaces},}\ }\href@noop {}
  {\bibfield  {journal} {\bibinfo  {journal} {Phys. Chem. Chem. Phys.}\
  }\textbf {\bibinfo {volume} {13}},\ \bibinfo {pages} {12447--12453} (\bibinfo
  {year} {2011})}\BibitemShut {NoStop}%
\bibitem [{\citenamefont {Chen}\ \emph {et~al.}(2016)\citenamefont {Chen},
  \citenamefont {Cheng}, \citenamefont {Xu}, \citenamefont {Meng},
  \citenamefont {Yuan}, \citenamefont {Liu},\ and\ \citenamefont
  {Liu}}]{chen_scirep_2016}%
  \BibitemOpen
  \bibfield  {author} {\bibinfo {author} {\bibfnamefont {L.}~\bibnamefont
  {Chen}}, \bibinfo {author} {\bibfnamefont {Z.}~\bibnamefont {Cheng}},
  \bibinfo {author} {\bibfnamefont {W.}~\bibnamefont {Xu}}, \bibinfo {author}
  {\bibfnamefont {X.}~\bibnamefont {Meng}}, \bibinfo {author} {\bibfnamefont
  {G.}~\bibnamefont {Yuan}}, \bibinfo {author} {\bibfnamefont {J.}~\bibnamefont
  {Liu}}, \ and\ \bibinfo {author} {\bibfnamefont {Z.}~\bibnamefont {Liu}},\
  }\bibfield  {title} {\enquote {\bibinfo {title} {{Electrical and mechanical
  switching of ferroelectric polarization in the 70 nm BiFeO$_3$ film}},}\
  }\href@noop {} {\bibfield  {journal} {\bibinfo  {journal} {Sci. Rep.}\
  }\textbf {\bibinfo {volume} {6}},\ \bibinfo {pages} {19092} (\bibinfo {year}
  {2016})}\BibitemShut {NoStop}%
\bibitem [{\citenamefont {Gattinoni}, \citenamefont {Efe},\ and\ \citenamefont
  {Spaldin}(2020)}]{prequel}%
  \BibitemOpen
  \bibfield  {author} {\bibinfo {author} {\bibfnamefont {C.}~\bibnamefont
  {Gattinoni}}, \bibinfo {author} {\bibfnamefont {I.}~\bibnamefont {Efe}}, \
  and\ \bibinfo {author} {\bibfnamefont {N.}~\bibnamefont {Spaldin}},\
  }\bibfield  {title} {\enquote {\bibinfo {title} {The half polarization
  quantum in {BiFeO}$_3$ and its consequences for thin films and
  heterostructures},}\ }\href@noop {} {\bibfield  {journal} {\bibinfo
  {journal} {in preparation}\ } (\bibinfo {year} {2020})}\BibitemShut {NoStop}%
\end{thebibliography}%


\begin{thebibliography}{4}%
\makeatletter
\providecommand \@ifxundefined [1]{%
 \@ifx{#1\undefined}
}%
\providecommand \@ifnum [1]{%
 \ifnum #1\expandafter \@firstoftwo
 \else \expandafter \@secondoftwo
 \fi
}%
\providecommand \@ifx [1]{%
 \ifx #1\expandafter \@firstoftwo
 \else \expandafter \@secondoftwo
 \fi
}%
\providecommand \natexlab [1]{#1}%
\providecommand \enquote  [1]{``#1''}%
\providecommand \bibnamefont  [1]{#1}%
\providecommand \bibfnamefont [1]{#1}%
\providecommand \citenamefont [1]{#1}%
\providecommand \href@noop [0]{\@secondoftwo}%
\providecommand \href [0]{\begingroup \@sanitize@url \@href}%
\providecommand \@href[1]{\@@startlink{#1}\@@href}%
\providecommand \@@href[1]{\endgroup#1\@@endlink}%
\providecommand \@sanitize@url [0]{\catcode `\\12\catcode `\$12\catcode
  `\&12\catcode `\#12\catcode `\^12\catcode `\_12\catcode `\%12\relax}%
\providecommand \@@startlink[1]{}%
\providecommand \@@endlink[0]{}%
\providecommand \url  [0]{\begingroup\@sanitize@url \@url }%
\providecommand \@url [1]{\endgroup\@href {#1}{\urlprefix }}%
\providecommand \urlprefix  [0]{URL }%
\providecommand \Eprint [0]{\href }%
\providecommand \doibase [0]{http://dx.doi.org/}%
\providecommand \selectlanguage [0]{\@gobble}%
\providecommand \bibinfo  [0]{\@secondoftwo}%
\providecommand \bibfield  [0]{\@secondoftwo}%
\providecommand \translation [1]{[#1]}%
\providecommand \BibitemOpen [0]{}%
\providecommand \bibitemStop [0]{}%
\providecommand \bibitemNoStop [0]{.\EOS\space}%
\providecommand \EOS [0]{\spacefactor3000\relax}%
\providecommand \BibitemShut  [1]{\csname bibitem#1\endcsname}%
\let\auto@bib@innerbib\@empty
\bibitem [{\citenamefont {Guhl}, \citenamefont {Miller},\ and\ \citenamefont
  {Reuter}(2010)}]{guhl_prb_2010}%
  \BibitemOpen
  \bibfield  {author} {\bibinfo {author} {\bibfnamefont {H.}~\bibnamefont
  {Guhl}}, \bibinfo {author} {\bibfnamefont {W.}~\bibnamefont {Miller}}, \ and\
  \bibinfo {author} {\bibfnamefont {K.}~\bibnamefont {Reuter}},\ }\bibfield
  {title} {\enquote {\bibinfo {title} {Water adsorption and dissociation on
  {SrTiO}$_{3}(001)$ revisited: A density functional theory study},}\ }\href
  {\doibase 10.1103/PhysRevB.81.155455} {\bibfield  {journal} {\bibinfo
  {journal} {Phys. Rev. B}\ }\textbf {\bibinfo {volume} {81}},\ \bibinfo
  {pages} {155455} (\bibinfo {year} {2010})}\BibitemShut {NoStop}%
\bibitem [{\citenamefont {Halwidl}\ \emph {et~al.}(2016)\citenamefont
  {Halwidl}, \citenamefont {Stöger}, \citenamefont {Mayr-Schmölzer},
  \citenamefont {Pavelec}, \citenamefont {Fobes}, \citenamefont {Peng},
  \citenamefont {Mao}, \citenamefont {Parkinson}, \citenamefont {Schmid},
  \citenamefont {Mittendorfer}, \citenamefont {Redinger},\ and\ \citenamefont
  {Diebold}}]{halwidl_2016}%
  \BibitemOpen
  \bibfield  {author} {\bibinfo {author} {\bibfnamefont {D.}~\bibnamefont
  {Halwidl}}, \bibinfo {author} {\bibfnamefont {B.}~\bibnamefont {Stöger}},
  \bibinfo {author} {\bibfnamefont {W.}~\bibnamefont {Mayr-Schmölzer}},
  \bibinfo {author} {\bibfnamefont {J.}~\bibnamefont {Pavelec}}, \bibinfo
  {author} {\bibfnamefont {D.}~\bibnamefont {Fobes}}, \bibinfo {author}
  {\bibfnamefont {J.}~\bibnamefont {Peng}}, \bibinfo {author} {\bibfnamefont
  {Z.}~\bibnamefont {Mao}}, \bibinfo {author} {\bibfnamefont {G.~S.}\
  \bibnamefont {Parkinson}}, \bibinfo {author} {\bibfnamefont {M.}~\bibnamefont
  {Schmid}}, \bibinfo {author} {\bibfnamefont {F.}~\bibnamefont
  {Mittendorfer}}, \bibinfo {author} {\bibfnamefont {J.}~\bibnamefont
  {Redinger}}, \ and\ \bibinfo {author} {\bibfnamefont {U.}~\bibnamefont
  {Diebold}},\ }\bibfield  {title} {\enquote {\bibinfo {title} {Adsorption of
  water at the {SrO} surface of ruthenates},}\ }\href@noop {} {\bibfield
  {journal} {\bibinfo  {journal} {Nature Mat.}\ }\textbf {\bibinfo {volume}
  {15}},\ \bibinfo {pages} {450--455} (\bibinfo {year} {2016})}\BibitemShut
  {NoStop}%
\bibitem [{\citenamefont {Evarestov}, \citenamefont {Bandura},\ and\
  \citenamefont {Alexandrov}(2007)}]{evarestov_2007}%
  \BibitemOpen
  \bibfield  {author} {\bibinfo {author} {\bibfnamefont {R.~A.}\ \bibnamefont
  {Evarestov}}, \bibinfo {author} {\bibfnamefont {A.~V.}\ \bibnamefont
  {Bandura}}, \ and\ \bibinfo {author} {\bibfnamefont {V.~E.}\ \bibnamefont
  {Alexandrov}},\ }\bibfield  {title} {\enquote {\bibinfo {title} {Adsorption
  of water on (001) surface of {SrTiO$_3$} and {SrZrO$_3$} cubic perovskites:
  {Hybrid} {HF}-{DFT} {LCAO} calculations},}\ }\href@noop {} {\bibfield
  {journal} {\bibinfo  {journal} {Surf. Sci.}\ }\textbf {\bibinfo {volume}
  {601}},\ \bibinfo {pages} {1844--1856} (\bibinfo {year} {2007})}\BibitemShut
  {NoStop}%
\bibitem [{\citenamefont {Bandura}, \citenamefont {Evarestov},\ and\
  \citenamefont {Kuruch}(2010)}]{bandura_2010}%
  \BibitemOpen
  \bibfield  {author} {\bibinfo {author} {\bibfnamefont {A.~V.}\ \bibnamefont
  {Bandura}}, \bibinfo {author} {\bibfnamefont {R.~A.}\ \bibnamefont
  {Evarestov}}, \ and\ \bibinfo {author} {\bibfnamefont {D.~D.}\ \bibnamefont
  {Kuruch}},\ }\bibfield  {title} {\enquote {\bibinfo {title} {Hybrid
  {HF}-{DFT} modeling of monolayer water adsorption on (001) surface of cubic
  {BaHfO$_3$} and {BaZrO$_3$} crystals},}\ }\href@noop {} {\bibfield  {journal}
  {\bibinfo  {journal} {Surf. Sci.}\ }\textbf {\bibinfo {volume} {604}},\
  \bibinfo {pages} {1591--1597} (\bibinfo {year} {2010})}\BibitemShut {NoStop}%
\end{thebibliography}%

\end{document}


\preprint{AIP/123-QED}

\title[Supplementary information]
  {Supplementary information}
%



\author{Ipek Efe}
\author{Nicola A. Spaldin }%
\author{Chiara Gattinoni}
\email{chiara.gattinoni@mat.ethz.ch}
\affiliation{ 
Materials Theory, Department of Materials, ETH Z{\"u}rich, Wolfgang-Pauli-Strasse 27, 8093, Z{\"u}rich, Switzerland}

\date{\today}

\maketitle

\section{Convergence test for slab thickness}

Calculations with the most favorable adsorption sites are carried out for the slab thicknesses of 2--6 unit cells (u.c.) for a water molecule on the charge-compensated slab of Fig. 1b.
All atoms were allowed to relax.
On the BiO surface the adsorption energy is converged for 4 u.c.
On the FeO$_2$ surface, the absolute values of the adsorption energies are still varying for a slab thickness of 6 u.c., however relative energies between the intact and dissociated structure are already converged at 4 u.c.
\begin{table}[h]
\begin{tabular}{|c | c  c | c  c|}
 \hline
 \multicolumn{5}{|c|}{Adsorption Energies (eV)} \\
 \hline
  & \multicolumn {2} {|c|} {\ce{FeO2} Surface} & \multicolumn {2} {|c|} {\ce{BiO} Surface} \\
 \hline
  Thickness & Intact & Dissociated & Intact & Dissociated \\
 \hline
 2 u.c.   & -0.788 & -0.424 & -0.664 & -0.617 \\
 \hline
 3 u.c.   & -0.593 & -0.560 & -0.456 & -0.425 \\
 \hline
 4 u.c.   & -0.804 & -0.451 & -0.672 & -0.642 \\
 \hline
 5 u.c.   & -0.787 & -0.407 & -0.654 & -0.624 \\
 \hline
 6 u.c.   & -0.784 & -0.428 &  & -0.607 \\
 \hline
\end{tabular}
\caption{ Adsorption energy of a water molecule on the compensated surfaces of Fig. 1b with slabs of varying thickness.}
\label{table:2}
\end{table}

\newpage

\section{Water on ``unhappy'' slab}

\begin{figure}[h]
\centering
 \includegraphics[width=0.8\textwidth]{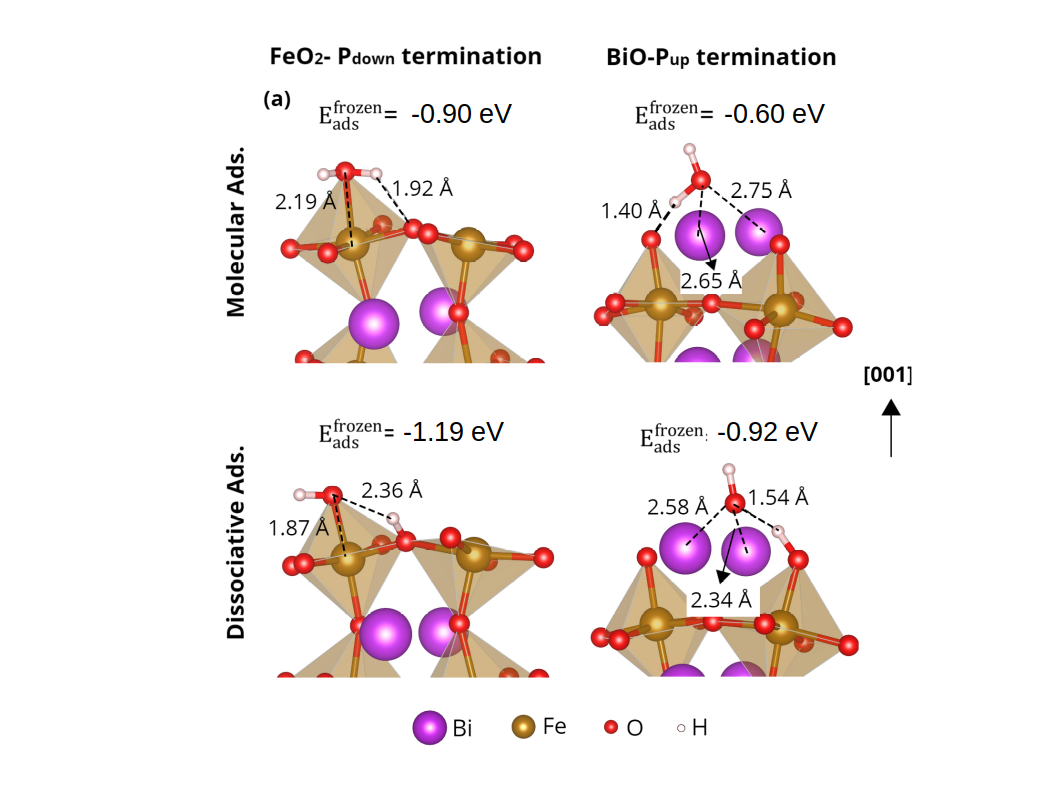}
 \caption{Most favorable adsorption geometries and the corresponding adsorption energies for a water molecule on the surfaces of frozen unhappy slab.} 
 \label{fig:bfo_si_unhappy}
\end{figure}

Structure of a single water molecule adsorbed on an ``unhappy'' frozen substrate, Fig.~\ref{fig:bfo_si_unhappy}.

\newpage

\section{Comparison of water adsorption behavior on perovskites}

\begin{table}[h]
\begin{tabular}{|c | c | c | c | c | c | c |}
 \hline
  \textbf{Reference} & \textbf{Material} & \makecell{\textbf{Lattice}\\\textbf{constant ({\AA})}} & \textbf{H$_2$O on AO} & \textbf{H$_2$O on BO$_2$} & \textbf{Charged layers} & \textbf{DFT Functional} \\
 \hline
 \makecell{happy\\system} & \ce{BiFeO3} & 3.95 & Molecular & Molecular & Yes & optB86b-vdW \\
 \hline
 \makecell{unhappy\\system} & \ce{BiFeO3} & 3.95 & Dissociated & Dissociated & Yes & optB86b-vdW \\
 \hline
 ~\cite{guhl_prb_2010} & \ce{SrTiO3} & 3.94 & Dissociated & \makecell{ Dissociated/\\Molecular} & No & GGA-PBE \\
 \hline
 ~\cite{halwidl_2016} & \ce{Sr2RuO4} & 3.9 & Dissociated & N/A & No & opt86-vdW \\
 \hline
 ~\cite{evarestov_2007} & \ce{SrTiO3} & 3.94 & \makecell{Equal\\preference} & Molecular & No & B3LYP XC \\
 \hline
 ~\cite{evarestov_2007} & \ce{SrZrO3} & 4.19 & Dissociated & Molecular & No & B3LYP XC \\
 \hline
 ~\cite{bandura_2010} & \makecell{\ce{BaHfO3} \\ \ce{BaZrO3}}  & 4.19 & Dissociated & Dissociated & No & PBE0-XC \\
 \hline
\end{tabular}
\caption{Comparison of the adsorption modes on different systems and their properties in the literature and this work}
\label{comparison}
\end{table}

\bibliography{references_bfo_water}